\documentclass[12pt]{article}
\usepackage{amsthm,amsmath,amsfonts,amssymb,cases}
\usepackage{xcolor}
\usepackage{comment}
\usepackage{enumerate}
\usepackage[colorlinks, linkcolor = black, citecolor = blue, filecolor = black, urlcolor = black]{hyperref}
\usepackage{cleveref}
\usepackage{tikz}
\usepackage{pgfplots}
\usepackage{hyperref}
\usepackage{lmodern}
\usetikzlibrary{decorations.pathmorphing}
\tikzset{snake it/.style={decorate, decoration=snake}}
\usepackage{float}
\usepackage[shortlabels]{enumitem}
\numberwithin{equation}{section}
\usepackage[style=alphabetic,giveninits=true,maxnames=4, backend=biber,doi=false,url=false,isbn=false,eprint=false]{biblatex}
\newcommand{\R}{{\mathord{\mathbb R}}}
\newcommand{\Z}{{\mathord{\mathbb Z}}}
\newcommand{\N}{{\mathord{\mathbb N}}}

\newcommand{\T}{{\mathord{\mathbb T}}}

\newcommand{\vertiii}[1]{{\left\vert\kern-0.25ex\left\vert\kern-0.25ex\left\vert #1 
		\right\vert\kern-0.25ex\right\vert\kern-0.25ex\right\vert}}

\newcommand{\ben}{\begin{displaymath}}
	\newcommand{\een}{\end{displaymath}}
\newcommand{\beqn}{\begin{equation}}
	\newcommand{\eeqn}{\end{equation}}
\newcommand{\beqna}{\begin{eqnarray*}}
	\newcommand{\eeqna}{\end{eqnarray*}}

\newcommand{\nn}{\nonumber} 

\renewcommand{\Im}{{\rm Im}}

\newtheorem{lemma}{Lemma}
\newtheorem{theorem}[lemma]{Theorem}
\newtheorem{remark}[lemma]{Remark}
\newtheorem{proposition}[lemma]{Proposition}
\newtheorem{corollary}[lemma]{Corollary}
\newtheorem{definition}[lemma]{Definition}

\numberwithin{equation}{section}
\numberwithin{lemma}{section}

\makeatletter
\newsavebox{\@brx}
\newcommand{\llangle}[1][]{\savebox{\@brx}{\(\m@th{#1\langle}\)}%
	\mathopen{\copy\@brx\kern-0.5\wd\@brx\usebox{\@brx}}}
\newcommand{\rrangle}[1][]{\savebox{\@brx}{\(\m@th{#1\rangle}\)}%
	\mathclose{\copy\@brx\kern-0.5\wd\@brx\usebox{\@brx}}}
\makeatother

\begin{document}
	\title{On Hölder Equicontinuity of the Integrated Density of States for Random Schr\"odinger Operators
on $\Z^d$}
	\author{\vspace{5pt}  Jan Fischer$^1$\footnote{E-mail: jan.fischer@uni-jena.de} \quad 
David Hasler$^1$\footnote{
			E-mail: david.hasler@uni-jena.de}  \\
		\vspace{-4pt} \small{$1.$ Department of Mathematics,
			Friedrich Schiller  University  Jena} \\ \small{Jena, Germany } }
	\date{}
	\maketitle
	
	\begin{abstract}
We study  a random Schr\"odinger operator $H_\lambda= H_0 + \lambda V$ on $\Z^d$
		consisting  of a translation invariant hopping matrix  $H_0$   and a random potential $V$, given by independently distributed single site potentials with 	absolutely continuous distributions. 
Assuming a regularity assumption for the hopping matrix $H_0$, 
	we show Hölder continuity 
		of  the integrated density of states of $H_\lambda$ up to exponents $2/3$,  independently of the disorder strength $\lambda$.
	\end{abstract}

	\section{Introduction}

	The random Schrödinger operators    we consider are of the type
	\begin{align*}
		H_{\lambda} = H_0 + \lambda V_\omega
	\end{align*}
	on $\ell^2(\mathbb{Z}^d)$ where the disorder strength $\lambda$ is non negative, cf. \cite{Anderson.1958,CarmonaLacroix}.
For these two operators we make the following general assumptions:
	\begin{enumerate}
		\item The operator $H_0$ is a bounded, selfadjoint and translation invariant operator, so the commutator satisfies $[H_0,S_\xi] = 0$ for every translation $S_\xi$ with $\xi\in\mathbb{Z}^d$ and $S_\xi \psi(x) = \psi(x+\xi)$ for all $\psi\in\ell^2(\mathbb{Z}^d).$
		\item The operator $V_\omega$ is a multiplication operator with $V_\omega\psi(x) = \omega_x\psi(x)$ where $\lbrace \omega_x \rbrace_{x\in\mathbb{Z}^d}$ are independent and identically distributed real valued random variables with a common bounded density function $\rho.$
	\end{enumerate}
Let us denote by  $\mathcal{B}_\R$ the set of Borel measureable subsets of the real line.
	If $(\mathbb{R},\mathcal{B}_\mathbb{R}, P)$ is the measure space for $\omega_0$, then our  canonical probability space is
	$ (\Omega, \mathcal{A},\mathbf{P})  = \bigotimes_{i\in\mathbb{Z}^d}(\mathbb{R},\mathcal{B}_{\mathbb{R}}, P)$.
	More explicitly we have $\Omega = \mathbb{R}^{\mathbb{Z}^d}$ and $\mathcal{A}$ is the $\sigma$-algebra generated by cylindrical sets, i.e., $A = \lbrace \omega \in \mathbb{R}^{\mathbb{Z}^d} \,|\, \omega_{i_1} \in A_1, \dots , \omega_{i_n}\in A_n\rbrace$ with $A_1, \dots A_n \in\mathcal{B}_{\mathbb{R}}$ and for some $i_1,\dots i_n\in\mathbb{Z}^d$, $n\in\mathbb{N}$. The probability measure ${\bf P}$ is given by
	$\mathrm{d}{\bf P}(\omega) = \prod_{i\in\mathbb{Z}^d}\rho(\omega_i)\mathrm{d}\omega_i$.
Expectations with respect to ${\bf P}$ will be denoted by ${\bf E}$. 

Define by $\mathbb{T}^d := \R^d /  (2 \pi \mathbb{Z})^d $ the $d$-dimensional torus.
	The operator $H_0$ can be diagonalized via Fourier transform.
More explicitly, there exists  a measurable, bounded,  and $2\pi$--periodic function   $\mathfrak{e} : \T^d \to \R$
such that 
	\begin{align}
		H_0 = \sum_{\xi\in\mathbb{Z}^d} \check{\mathfrak{e}}(\xi)S_\xi ,
	\end{align}
where  $ \check{\mathfrak{e}}$  is the inverse Fourier transform of $\mathfrak{e}$ 
	\begin{align*}
		\check{\mathfrak{e}}(\xi)= \int_{\T^d} \mathfrak{e}(k) e^{-ik\cdot \xi}d\widehat{k}, \qquad  \xi \in \Z^d  ,
	\end{align*}
where we introduced the normalized Lebesgue measure 
	\begin{align}  \label{normalizedmeasure} 
  d\widehat{k}= \frac{dk}{(2\pi)^d}.
	\end{align}
	This sum is understood as a Fourier series and not as a absolute convergent sum.
Observe that for $\psi_k(x) = e^{ i k \cdot  x } $ with $k \in \T^d$ it follows from the definitions that 
\begin{align} 
 H_0 \psi_k =   \sum_{\xi\in\mathbb{Z}^d} \check{\mathfrak{e}}(\xi)S_\xi \psi_k = \sum_{\xi\in\mathbb{Z}^d} \check{\mathfrak{e}}(\xi)e^{  i k \cdot\xi}   \psi_k = \mathfrak{e}(k) \psi_k , 
 \label{diagfreeham}
\end{align} 
where in the last identity we made use of  the Fourier transform. 
Thus  in momentum space $H_0$ is just the multiplication with $\mathfrak{e}$ and by the spectral theorem we can define $f(H_0)$ for bounded measurable functions.
	\begin{definition}
		We define the {\bf density of states measure}  of $H_\lambda$ by 
		\begin{align}
			N_\lambda(B)=\mathbf{E}\langle\delta_0,1_B(H_\lambda)\delta_0\rangle
		\end{align}	for all $B\in \mathcal{B}_{\mathbb{R}}$.
	\end{definition}
Note that in the free case, $\lambda =0$,    
		\begin{align} \label{freeDOS}
				N_0(B)=\langle\delta_0,1_B(H_0)\delta_0\rangle  , \qquad B \in \mathcal{B}_{\mathbb{R}}.
		\end{align}
Our results will depend on  local information of the density of states of $H_0$.  More specifically,
we introduce the following definition. 
	\begin{definition} \label{def:regularpoint} 
		A  point $E \in \R$ is called a \textbf{regular point} for the operator $H_0$ if there exists  a $\delta_0 \in (0,1)$ and a constant $\Gamma(E)$ such that
		\begin{align}\label{regularpoint1} 
			N_0([E-\delta,E+\delta]) \leq \Gamma(E)\delta 
		\end{align}
		for all $\delta \in [0,\delta_0]$. \\
		More generally,  a point $E \in \R$ is called a \textbf{regular point of order $\alpha \in (0,1]$} if there exists  a $\delta_0 \in (0,1)$ and a constant $\Gamma(E,\alpha)$ such that
		\begin{align} \label{regularpoint} 
			N_0([E-\delta,E+\delta]) \leq \Gamma(E,\alpha)\delta^\alpha.
		\end{align}
		for all $\delta\in [0, \delta_0]$.
	\end{definition}
In this paper we are interested in estimates of the following form.  Let  $\lambda >  0$,
Given $E \in \R$, we ask whether there are  $\beta \geq 0$  and $\delta_0 > 0$ such that 
 		\begin{align} \label{int:goal} 
			N_\lambda([E-\delta,E+\delta]) \le C(E,\rho,\beta) \delta^{ \beta}  , \quad \text{ for all }   \delta \in [0,\delta_0 ] .
		\end{align}
  If $\rho$ is bounded, we know by the Wegner estimate  \cite{Wegner.1981} that 
 		\begin{align} \label{int:wegner} 
			N_\lambda([E-\delta,E+\delta]) \le \frac{ \|\rho\|_\infty}{\lambda} 2\delta  , \quad \text{ for all }   \delta \geq 0 
		\end{align}
and $\lambda > 0$.  This is a good estimate for large $\lambda$.
but unfortunately  the right hand side of \eqref{int:wegner} diverges as $\lambda \downarrow 0$. 
On the other hand, for $\lambda =0$, we have the free density measure which can be explicitly computed
in terms of the energy function $\mathfrak{e}$ (i.e. the push forward measure $\mathfrak{e}_* (\hat{dk})$).
Thus we ask,  what happens in the intermediate regime, when $\lambda$ tends to zero.
It is natural to conjecture (cf. \cite{Schenker.2004}) that  there  exists a constant uniformly bounded in $\lambda$
 such that  \eqref{int:goal}  holds for $\beta=1$. In fact this conjecture has been shown for the Cauchy distribution \cite{Loyd.1969,CarmonaLacroix}, and on  the Bethe lattice or trees \cite{AcostaKlein1992,AizenmanWarzel.2011}.

This question was addressed in  \cite{Schenker.2004}, where the weaker statement  for $\beta < 1/2$ could be shown. 
If $\mathbf{E} |\omega_0|^n < \infty$ for some $n \in (2,\infty)$ and $E$ is a regular point 
of $\mathfrak{e}$, then there exists
a constant  $C(E,\rho)$ uniformly  bounded in $\lambda \geq 0$ such that 
Ineq.   \eqref{int:goal}  holds   for $\beta = \frac{1}{2}(1-\frac{1}{2n+1} )$ 
and  all $\delta \geq 0$. If all moments exist,  $n$ can be chosen arbitrarily large and one 
obtains \eqref{int:goal} uniformly in $\lambda \geq 0$ for any  $\beta < 1/2$. 
More generally,  if $E$ is only a point of  regular order $\alpha $ for $\mathfrak{e}$, then 
  there exists
a constant  $C(E,\rho,\alpha)$ uniformly  in $\lambda \geq 0$ such that 
Ineq.   \eqref{int:goal}  holds   for $\beta = \frac{\alpha}{\alpha+1}(1-\frac{1}{\frac{\alpha+1}{\alpha} n+1} )$ 
and  all $\delta \geq 0$.  In  \cite{Schenker.2004} a novel method was introduced to show 
the results, but it is noted, that the methods in \cite{CombesHislopKlopp.2003} could alternatively be used 
to obtain an analogous result. 

In the present paper, we  use ideas   from \cite{CombesHislopKlopp.2003} and prove  results similar to   \cite{Schenker.2004}. Moreover, combining these estimates with  expansion methods, similar to \cite{ErdosSalmhoferYau.2007a,ErdosSalmhoferYau.2007b,ErdosSalmhoferYau.2008a,HaslerKoberstein.2024-2,FischerHaslerKoberstein.2026},  we can improve the results in  \cite{Schenker.2004}  
up to  $\beta < 2/3$. Which is somewhat  closer to the conjectured
value  $\beta =1$.

We  show that the larger the order $n$ of the moments    $\mathbf{E}[|\omega_0|^n]$ are finite,
the larger $\beta$ can be chosen. 
However, there are interesting distributions  where not all moments 
exist\footnote{For example  for the  Cauchy distribution   $\mathbf{E}[|\omega_0|^p]$ diverges if $p \geq 1$.} and thus it is of interest, to establish estimates requiring only finiteness of the, say, first moments. 
We thus present  our results  in increasing order of  the  required  moments in the  assumptions.
\\
$\bullet$ Let $\mathbf{E}[|\omega_0|^n] < \infty$ for some $1\le n<\infty$ and let $E$ be a regular point
of order $\alpha$. Then there exists a constant 
 $C(E,\rho,n,\alpha)<\infty$ such that
		\begin{align} \label{firstresultsummary}
			N_\lambda([E-\delta,E+\delta]) \leq C(E,\rho,n,\alpha)\delta^{\frac{\alpha}{\alpha + 1 +\alpha/n}}.
		\end{align}
		for all $\lambda,\delta\ge0$.  
This is shown in Theorem \ref{regpoint alpha} for general $\alpha$. We note in complement  to  \cite{Schenker.2004} we only require $n \ge 1$. As $n \to \infty$ the exponent $\beta$ in \eqref{firstresultsummary} for $\alpha=1$ also converges  to  $1/2$ from below as in \cite{Schenker.2004}. The proof given in  Section	\ref{sec:FirstMomentEstimates}  uses the idea from \cite{CombesHislopKlopp.2003} and only require a single application of the second resolvent identity.
\\ 
$\bullet$ Let $E$ be a regular point. If  $\mathbf{E}[\omega_0^2]<\infty$ and 
$\beta = \frac{1}{2}$, or if $\mathbf{E}[\omega_0^4]<\infty$, $\mathbf{E}[\omega_0]=0$, and $\beta = \frac{4}{7}$, then there 
exists a constant $C(\rho, E)$  such that  
		\begin{align}
			N_\lambda([E-\delta,E+\delta])\leq C(\rho, E)\delta^{\beta} ,
		\end{align}
		for all $\lambda ,\delta \geq 0$, cf. Theorems \ref{Improved1} and   \ref{Improved}, respectively.
In the proofs, presented in Section \ref{sec:lowermoment}, we apply the second resolvent identity two respectively  four times. 
\\
$\bullet$ Let $E$ be a regular point and $\mathbf{E}[\omega_0]=0$.
If $ \mathbf{E}[\omega_0^{2n}]<\infty$ for some $n \geq 3$, then
 there exists a finite constant $C(E,\rho, n)$ such that 
		\begin{align} \label{int:finalresult} 
			N_\lambda([E-\delta,E+\delta]) \le C(E,\rho, n) \delta^{\frac{2n}{3n+1}}\left(1+\frac{2n}{3n+1} |\ln \delta|\right)^{\frac{2n}{2n+1}}
		\end{align}
		for all $\lambda,\delta\ge0$. This result is stated in Theorem \ref{thm:genorder}.
We note that as  $n \to \infty$ the right hand side in \eqref{int:finalresult} 
can be estimated in terms of  $ C(E,\rho, n)  \delta^{\beta_n}$, where $\beta_n$ converges to $2/3$ from below, cf. Corollary  \ref{corollarymain}.  The idea of the proof  given  in Section \ref{sec:HigherMomentEstimates} uses the second resolvent identity   $2n$ times  and estimates on  the resulting 
expressions, which are shown  in Section \ref{sec:feynmanbounds}, using ideas from 
\cite{ErdosSalmhoferYau.2007a,ErdosSalmhoferYau.2007b,ErdosSalmhoferYau.2008a,HaslerKoberstein.2024-2,FischerHaslerKoberstein.2026}.

 Related  results on  uniform H\"older continuity have been obtained  in the special case of Gaussian random variables   in the Master's thesis of one of the authors \cite{FischerMaster} 
with  exponents  up to $4/7$ 
and for Poisson distributed random Schr\"odinger in the continuum
\cite{FischerHaslerKoberstein.2026} with exponents up to $2/3$.
All proofs use an  interpolation argument involving  Wegner's estimate, which is recalled in  Appendix \ref{sec:wegner}.
In the following  Section \ref{sec:FiniteVolumeApproximation}, we show how 
operators  acting on  the  $\ell^2$--space of a finite subset of the lattice can be used to obtain bounds on the infinite volume integrated density of states. 
 In Appendix \ref{sec:integralbounds} we give integral bounds on the integrated density of states for the kinetic energy, which 
are used in the proofs of Theorems   \ref{Improved1},  \ref{Improved}, and  \ref{thm:genorder}.

For the proofs  we will work with the   Schatten class trace norms which 
are defined  for a compact operator 
$K$ in a  Hilbert space   by 
\begin{align}
\| K \|_p =  ( {\rm tr}  |K|^p  )^{1/p} ,  \quad p \in [1,\infty) . 
\end{align}

	\section{Finite Volume Approximation}
\label{sec:FiniteVolumeApproximation}

In order to estimate the density of states measure $N_\lambda$, we approximate it by  density of states measures 
for operators acting on finite  subsets of $\Z^d$.  
In the finite case, we can calculate the  density of states by means of  the trace  and explicit calculations  using  the discrete Fourier transform,
as  the resulting operators are just matrices acting on a finite dimensional subspace.
We will 
show  in Propositions   \ref{prop:dosinfinite}  and \ref{prop:limids}  that the corresponding finite volume density of states measure converges weakly to $N_\lambda$.

We define for $L\in \mathbb{N}$ the set  $$\Lambda_L = \mathbb{Z}^d \cap [-L/2, L/2)^d$$ 
and the  dual $$\hat{\Lambda}_L = [-\pi,\pi)^d \cap (2\pi\mathbb{Z}/L)^d.$$
	This will be the index set for the momentum space.  Note that  $|\hat{\Lambda}_L|= |\Lambda_L| = L^d $. 
	For each $k\in\hat{\Lambda}_L$ define the cell 
	\begin{align*}
		Q_{L,k} =k +[-\pi/L,\pi/L)^d
	\end{align*}
	with addition understood  modulo $2\pi$ in each coordinate. 
Observe that $(Q_{L,k})_{k \in \hat{\Lambda}_L}$ is a partition of $\T^d$.  
Define for each $k\in\hat{\Lambda}_L$ the average of the function on the cell $Q_{L,k}$ as
	\begin{align*}
		\mathfrak{e}_L(k)=|\Lambda_L| \int_{Q_{L,k}}\mathfrak{e}(q)d\widehat{q}
	\end{align*}
	with $d\widehat{q}$ the normalized measure    \eqref{normalizedmeasure}, for which $\int_{Q_{L,k}} 1 d \widehat{q} = |\Lambda_L|^{-1}$. This implies $\mathfrak{e}_L(k)\in \mathbb{R}$ and $|\mathfrak{e}_L(k)|\le \| \mathfrak{e}\|_{L^\infty}$. Furthermore,  define  $\mathfrak{e}^{(L)} : \T^d \to \R$ by 
	\begin{align*}
		\mathfrak{e}^{(L)}(q)= \mathfrak{e}_L(k), \quad  \text{  if  } q\in Q_{L,k}  \quad \text{ for } k \in  \hat{\Lambda}_L , 
	\end{align*}
	which can be viewed as a piecewise  approximation of $\mathfrak{e}$.
	
	We get a new finite dimensional Hilbert space $\ell^2(\Lambda_L)$ and define on this the operator 
	\begin{align*}
		H_\lambda^L = H_0^L+\lambda V_L
	\end{align*}
	with $V_L$ is just the multiplication with $V_L \psi(x) = \omega_x \psi(x)$ for $x\in \Lambda_L$ and $H_0^L$ is the finite volume operator with periodic boundary condition, that is,
	$$H_0^L\varphi(x) = \sum_{\xi\in\Lambda_L}\check{\mathfrak{e}}_L(\xi)\tilde\varphi(x+\xi) $$
	with $\tilde\varphi$ being the periodic extension to $\mathbb{Z}^d$ and 
	\begin{align*} 
		\check{\mathfrak{e}}_L(\xi) = |\Lambda_L|^{-1}\sum_{k\in\hat{\Lambda}_L}\mathfrak{e}_L(k)e^{-ik\cdot \xi}.
	\end{align*}
	We can diagonalize $H_0^L$ on $\ell^2(\Lambda_L)$ with respect to the orthonormal basis
	\begin{align} \label{eigenbasisfinite} 
		\varphi_k(x) = {|\Lambda_L|^{-1/2}}e^{ik\cdot x},\quad k\in\hat{\Lambda}_L, \quad x \in  \Lambda_L  ,
	\end{align}
	such that
	\begin{align} \label{eq:eigenvalueeqH0} 
		H_0^L \varphi_k =  \sum_{\xi\in\Lambda_L}\check{\mathfrak{e}}_L(\xi) e^{ik\cdot\xi} \varphi_k= \mathfrak{e}_L(k)\varphi_k.
	\end{align}
	Here we use $\sum_{\xi\in\Lambda_L}e^{i(k-q)\cdot \xi}= |\Lambda_L| \delta_{k,q}$. 
	\begin{lemma} \label{convergenceoffns} 
		We have 
		\begin{align*}
			\|\mathfrak{e}^{(L)}-\mathfrak{e}\|_{L^1}\to0,\quad \text{as } L\to\infty.
		\end{align*}
	\end{lemma}
	\begin{proof}
 	We use a standard density argument. Continuous functions are dense in $L^1(\T^d)$. Thus for any $\varepsilon>0$, there exists a $g\in C(\T^d)$, such that
	\begin{align*}
		\|\mathfrak{e}-g\|_{L^1}<\varepsilon.
	\end{align*}
	Hence
	\begin{align*}
		\|\mathfrak{e}^{(L)}-\mathfrak{e}\|_{L^1} &\le \|(\mathfrak{e}-g)^{(L)}\|_{L^1}+\|g^{(L)}-g\|_{L^1}+\|g-\mathfrak{e}\|_{L^1}\\
		&\le \|\mathfrak{e}-g\|_{L^1}+\|g^{(L)}-g\|_{L^1}+ \varepsilon \\
		&\le 2\varepsilon+\|g^{(L)}-g\|_{L^1}
	\end{align*}
	In the second inequality we used the Jensen inequality and the density argument. Since $g$ is continuous, we have $\|g^{(L)}-g\|_{L^\infty}\to0$ as $L\to \infty$. Thus the claim follows by letting $L\to\infty$ and $\varepsilon\downarrow0.$
	\end{proof} 
	This in particular implies weak convergence of the measure 
	\begin{align} \label{finitedospushforwards}
		N_0^L(B) := |\Lambda_L|^{-1}\mathrm{tr}1_B(H_0^L)= \int_{\T^d} 1_B(\mathfrak{e}^{(L)}(q))d\widehat{q} , \qquad B \in \mathcal{B}_\R  ,
	\end{align}
where the second equality follows from \eqref{eq:eigenvalueeqH0},
	to the density of states measure of the free operator
	\begin{align} \label{freedosispushforwarde} 
		N_0(B)= \langle \delta_0, 1_B(H_0)\delta_0\rangle = \int_{\T^d} 1_B(\mathfrak{e}(k))d\widehat{k} ,  \qquad B \in \mathcal{B}_\R , 
	\end{align}
where the first equality is simply the definition, cf. \eqref{freeDOS},  and the second equality follows from  \eqref{diagfreeham}.
	\begin{proposition} \label{prop:dosinfinite} 
		Let $L\in\mathbb{N}$ and $H_\lambda^L$ be the finite volume operator with periodic boundary as above. For $\lambda > 0$ we define the finite volume measure by
		\begin{align*}
			N_\lambda^L(B)=|\Lambda_L|^{-1}\mathbf{E}\mathrm{tr}1_B(H_\lambda^L) , \qquad B \in \mathcal{B}_\R  . 
		\end{align*}
		Then for every $\lambda\ge 0$ the measure $N_\lambda^L$ converges weakly to $N_\lambda$. In particular, for every bounded and continuous function $f$ we have
		\begin{align*}
			\lim_{L\to\infty}|\Lambda_L|^{-1}\mathbf{E}\mathrm{tr}f(H_\lambda^L)=\mathbf{E}\langle\delta_0,f(H_\lambda)\delta_0\rangle.
		\end{align*}
	\end{proposition}
	\begin{proof}
		The case $\lambda=0$ was already shown. Let us assume $\lambda>0$. We first show strong convergence of the extended free finite volume operator. Define 
		\begin{align*}
			I_L(x)=|\Lambda_L|\int_{[-\pi/L,\pi/L)^d}e^{-iq\cdot x}d\widehat{q} , \quad x \in \Z^d .
		\end{align*}
		Applying the Fourier transform to $\mathfrak{e}^{(L)}$ results in
	\begin{align}\label{propL-0} 
		\int_{\T^d}\mathfrak{e}^{(L)}(q)e^{-iq\cdot x}d\widehat{q}=I_L(x)\check{\mathfrak{e}}_L(x) , \quad x \in \Z^d   .
	\end{align}
	By a direct calculation we have 
	\begin{align} \label{propL-1} 
		I_L(x)=\prod_{j=1}^d\frac{\sin(\pi x_j/L)}{\pi x_j /L}\to 1 , \quad x \in \Z^d  ,
	\end{align}
	as $L\to\infty$. Here we use the convention, that a factor with $x_j=0$ is equal to one. We know that $\|\mathfrak{e}^{(L)}-\mathfrak{e}\|_{L^1}\to 0$, hence for fixed $x \in \Z^d $
	\begin{align} \label{propL-2} 
			\int_{\T^d}\mathfrak{e}^{(L)}(q)e^{-iq\cdot x}d\widehat{q} \to 	\int_{\T^d}\mathfrak{e}(q)e^{-iq\cdot x}d\widehat{q} = \check{\mathfrak{e}}(x).
	\end{align}  
	Thus for every fixed $x\in \Z^d $ it follows from \eqref{propL-0}, \eqref{propL-1}, and  \eqref{propL-2} 
	\begin{align}
		\lim_{L\to\infty} \check{\mathfrak{e}}_L(x) 
= \check{\mathfrak{e}}(x) .  \label{fourier conv}
	\end{align}
	Now let 
	\begin{align*}
		\iota_L:\ell^2(\Lambda_L)\to\ell^2(\mathbb{Z}^d)
	\end{align*}
	be the extension by zero operator and define
	\begin{align*}
		T_L = \iota_L H_0^L \iota_L^\ast.
	\end{align*}
	Recall that the eigenvalues of $H_0^L$ are $\mathfrak{e}_L(k)$, thus
	\begin{align*}
		\|T_L\|=\|H_0^L\|\le \| \mathfrak{e}\|_{L^\infty}.
	\end{align*}
	So the operator $T_L$ is uniformly bounded.
	
	For fixed lattice sites $x,y\in\mathbb{Z}^d$, we have by definition 
	\begin{align*}
		\langle \delta_x,T_L\delta_y\rangle=\check{\mathfrak{e}}_L(y-x)
	\end{align*}
	for sufficiently large $L$ with $x,y\in\Lambda_L$. On the other hand we have 
	\begin{align*}
			\langle \delta_x,H_0\delta_y\rangle=\check{\mathfrak{e}}(y-x).
	\end{align*}
	Hence by \eqref{fourier conv}
	\begin{align}
			\langle \delta_x,T_L\delta_y\rangle \to 	\langle \delta_x,H_0\delta_y\rangle. \label{weak conv matrix}
	\end{align}
	
	We now want to show strong convergence. To do this we define the discrete Fourier transform of the square by
	\begin{align*}
		s_L(x)=|\Lambda_L|^{-1}\sum_{k\in\hat{\Lambda}_L}\mathfrak{e}_L(k)^2e^{-ik\cdot x} , \quad x \in \Z^d . 
	\end{align*}
	The same calculation as above gives
	\begin{align*}
		\int_{\T^d}(\mathfrak{e}^{(L)}(q))^2e^{-iq\cdot x}d\widehat{q}=s_L(x)I_L(x) , \quad x \in \Z^d . 
	\end{align*} 
	Since $|\mathfrak{e}^{(L)}(q)|\le \| \mathfrak{e}\|_{L^\infty}$ we have
	\begin{align}
		\|(\mathfrak{e}^{(L)})^2-\mathfrak{e}^2\|_{L^1}\le 2\|\mathfrak{e}\|_{L^\infty}\|\mathfrak{e}^{(L)}-\mathfrak{e}\|_{L^1} \to 0. \label{fourier conv 2}
	\end{align}
	Hence for every fixed $x \in\mathbb{Z}^d$
	\begin{align*}
		s_L(x) \to	\int_{\T^d}\mathfrak{e}(q)^2 e^{-iq\cdot x}d\widehat{q}.
	\end{align*}
	Notice that $\iota_L^\ast\iota_L$ is the identity on $\ell^2(\Lambda_L)$, thus
	\begin{align*}
		T_L^2=\iota_L(H_0^L)^2\iota_L^\ast.
	\end{align*}
	By \eqref{fourier conv 2} we have convergence of the matrix elements of $T_L^2$ to $H_0^2$. Now let $\psi$ be a function of finite support. From \eqref{weak conv matrix} we have for finitely supported $\varphi$
	\begin{align*}
		\langle\varphi,T_L\psi\rangle \to \langle\varphi,H_0\psi\rangle
	\end{align*}
	Since $\sup_L\|T_L\| <\infty$ and $H_0$ is bounded, we can extend it to any $\varphi\in\ell^2(\mathbb{Z}^d)$ by density of the finitely supported functions. Thus we have weak convergence $T_L\psi \rightharpoonup H_0\psi$. Furthermore
	\begin{align*}
		\| T_L\psi\|^2 = \langle\psi, T_L^2\psi\rangle \to \langle\psi, H_0^2\psi\rangle=\|H_0\psi\|^2,
	\end{align*}
	since the matrix elements converge and $\psi$ has finite support. Now weak convergence and convergence of norms gives
	\begin{align*}
		\lim_{L\to\infty}\| T_L\psi -H_0\psi\|_{\ell^2(\mathbb{Z}^d)}=0.
	\end{align*}
	Finally, by the uniform bound of $\|T_L\|$ and the density of the finitely supported functions we get strong convergence of $T_L$ to $H_0$.\\
	
	We now extend the random finite volume operator the the infinite one. For each realization $\omega$ define 
	\begin{align*}
		A_\omega^L= T_L +\lambda V_\omega,\qquad A_\omega=H_0+\lambda V_\omega.
	\end{align*}
	We decompose $\ell^2(\mathbb{Z}^d)=\ell^2(\Lambda_L)\oplus \ell^2(\Lambda_L^c)$. Since $0\in \Lambda_L$, we have
	\begin{align*}
		\langle\delta_0,(A_\omega^L-z)^{-1}\delta_0\rangle=\langle\delta_0,(H_\lambda^L-z)^{-1}\delta_0\rangle.
	\end{align*}
	By the resolvent identity we get
	\begin{align*}
		(A_\omega^L-z)^{-1}-(A_\omega-z)^{-1}=(A_\omega^L-z)^{-1}(H_0-T_L)(A_\omega-z)^{-1}.
	\end{align*}
	Then for every $\psi\in\ell^2(\mathbb{Z}^d)$ we get by the strong convergence
	\begin{align*}
		\|\left(  	(A_\omega^L-z)^{-1}-(A_\omega-z)^{-1}\right)\psi\| \le \frac{1}{|\Im z|} \|(H_0-T_L)(A_\omega-z)^{-1} \psi\| \to 0.
	\end{align*}
	Thus we also have strong convergence for $A_\omega^L$ to $A_\omega$, which is $H_\lambda$.\\
	
	To finish the proof, we show that the Stieltjes tranforms converges. For $z\in\mathbb{C}\setminus\mathbb{R}$ let
	\begin{align*}
		m_L(z)=\int \frac{dN_\lambda^L(t)}{t-z}.
	\end{align*}
	By definition we have 
	\begin{align*}
		m_L(z)=|\Lambda_L|^{-1}\mathbf{E}\mathrm{tr}(H_\lambda^L-z)^{-1}.
	\end{align*}
	Now we can use translation invariance on the finite torus, that is, for each $x\in \Lambda_L$ 
	\begin{align*}
		\mathbf{E}\langle \delta_x,(H_\lambda^L-z)^{-1}\delta_x\rangle = \mathbf{E}\langle \delta_0,(H_\lambda^L-z)^{-1}\delta_0\rangle.
	\end{align*}
	This holds because periodic translations commute with $H_0^L$ by definition and translating the potential permutes the i.i.d. random variables and leaves the joint law unchanged. Hence expanding the trace above results in
	\begin{align*}
		m_L(z)=\mathbf{E}\langle \delta_0,(H_\lambda^L-z)^{-1}\delta_0\rangle =\mathbf{E}\langle \delta_0,(A_\omega^L-z)^{-1}\delta_0\rangle.
	\end{align*}
	By the strong convergence, the integrand converges almost surely
	\begin{align*}
		\langle \delta_0,(A_\omega^L-z)^{-1}\delta_0\rangle \to \langle \delta_0,(A_\omega-z)^{-1}\delta_0\rangle=\langle \delta_0,(H_\lambda-z)^{-1}\delta_0\rangle.
	\end{align*}
	Since $|\langle \delta_0,(A_\omega^L-z)^{-1}\delta_0\rangle|\le |\mathrm{Im}z|^{-1}$, dominated convergence gives
	\begin{align*}
		\lim_{L\to\infty}m_L(z)= \mathbf{E}\langle \delta_0,(H_\lambda-z)^{-1}\delta_0\rangle = \int\frac{dN_\lambda(t)}{t-z}.
	\end{align*}
	The right hand side is the Stieltjes transform of the density of states measure $N_\lambda$. Now both measures are in fact probability measures and by the property of Stieltjes transforms, we get weak convergence of $N_\lambda^L$ to $N_\lambda$.
	\end{proof}

 Proposition \ref{prop:dosinfinite} 	together with the Wegner estimate we obtain the following proposition. 
	\begin{proposition} \label{prop:limids} 
		Let $\lambda>0$. Then for every bounded interval $I\subset \mathbb{R}$ we get
		\begin{align*}
			\lim_{L\to\infty}|\Lambda_L|^{-1}\mathbf{E}\mathrm{tr}1_I(H_\lambda^L)=\mathbf{E}\langle\delta_0,1_I(H_\lambda)\delta_0\rangle=N_\lambda(I).
		\end{align*}
	\end{proposition}
\begin{proof}  Proposition \ref{prop:limids}  follows since the Wegner estimate,  Theorem  \ref{thm:wegner}
in the appendix, implies $N_\lambda(\{x\})=0$ for every $x\in\mathbb{R}$ and then we can apply the weak convergence property
of  Proposition \ref{prop:dosinfinite} and use the Portmanteau theorem \cite[Theorem 13.16]{Klenke.2014}.
\end{proof}

	\begin{remark} \label{remeigenbasis}
		{\rm  For $\lambda = 0$ we do not have  this convergence, since the two boundary points of the interval $I$ could have non zero mass without additional information. But we still have a useful upper bound.  By  the weak convergence, see Lemma  \ref{convergenceoffns} and the discussion below,  
		 we find for any interval $I$ using the Portmanteau theorem \cite[Theorem 13.16]{Klenke.2014}.
			\begin{align}
				  \limsup_{L\to\infty} |\Lambda_L|^{-1} \sum_{k\in \hat{\Lambda}_L} 1_I(\mathfrak{e}_L(k))=
\limsup_{L\to\infty} N_0^L(I) \le N_0(\overline  I)= \int_{\T^d}  1_{\overline I}(\mathfrak{e}(k)) d \widehat{k}. 
			\end{align} 
		Note that for  the standard discrete Laplace operator, we do not have this problem\footnote{This follows,  since  the set of points in $\T^d$ where the gradient of $k \mapsto \sum_j \cos(k_j)$ vanishes
does not have positive measure.}.
		} 
	\end{remark} 
To estimate expressions of the form
	$$
	|\Lambda_L|^{-1} \sum_{s \in \hat{\Lambda}_L } F(\mathfrak{e}_L( s  ))  ,
	$$
we recall that by definition of $\mathfrak{e}^{(L)}$, Lemma  \ref{convergenceoffns}
and the discussion below, we have 
	\begin{align*} 
		|\Lambda_L|^{-1} \sum_{s \in \hat{\Lambda}_L } F(\mathfrak{e}_L( s  )) 
=  \int_{\T^d}  F(\mathfrak{e}^{(L)}(s)) d \widehat{s}  = \int F d N_0^L  \to  \int F d N_0 = \int_{\T^d}  F(\mathfrak{e}(s)) d \widehat{s}  
	\end{align*} 
for any   continuous  and bounded function $F$.  To apply this to functions involving characterstic functions of intervals, which are not continuous, 
we will use the following: 
	If  $f$ is a bounded and upper semicontinuous function,  then the Portmanteau theorem, cf. 
\cite[Corollary 8.2.5]{Bogachev2007Vol2}, gives
	\begin{align} \label{uppersemig} 
		\limsup_{L\to\infty}\int f dN_0^L \le \int fdN_0.
	\end{align}
	Hence it suffices to work with  integrable majorants, that are bounded and upper semicontinuous.
	
	\section{First Moment Estimates} 
\label{sec:FirstMomentEstimates}

In this section we show H\"older estimates uniform in the coupling strength similar to the ones obtained in  \cite{Schenker.2004}, but with a different method.
The method used here  is similar to the one  used in     \cite{CombesHislopKlopp.2003}, to obtain H\"older continuity for continuous random Schr\"odinger operators at fixed coupling strength. However in contrast to  \cite{CombesHislopKlopp.2003}, we only  expand the resolvent 
once, and we  work with the  Wegner estimate, which is available in the discrete setting, rather than the spectral shift function.

		\begin{theorem}\label{regpoint alpha}
		Assume that 
		$ \mathbf{E}[|\omega_0|^q] < \infty$
		for some $q \in [1,\infty)$,  or that  $\rho$ is compactly supported, in which case we set $q=\infty$. If $E$ is a regular point of order $\alpha$, then there is some finite constant $C(E,\rho,q,\alpha)<\infty$ such that
		\begin{align} \label{ineqforboundfirst} 
			N_\lambda([E-\delta,E+\delta]) \leq C(E,\rho,q,\alpha)\delta^{\frac{\alpha}{\alpha+1+\alpha/q}}.
		\end{align}
		for all $\lambda,\delta\ge 0$
	\end{theorem}

	\begin{proof}	The case $\lambda=0$  follows immdatiely from the definitions. So assume $\lambda>0$. 
If $\delta =0$ the left hand side is zero by  Wegner's estimate  \eqref{Wegner estimate} and so  \eqref{ineqforboundfirst}
follows in that case.  
Let $E$ be a regular point of order $\alpha$. Then  by Defintion \ref{def:regularpoint}  there exists a $\delta_0 \in ( 0,1)$ such that \eqref{regularpoint}  holds.  

To estimate the integrated density of states, we will use a  finite volume approximation by means of 
 Proposition \ref{prop:limids}.  For notational compactness we  will   suppress the $L$ label and write 
\begin{align}  \label{shorhandnot}
V, \, H_{\lambda} , \,   H_0 \quad \text{ for } \quad V_L, \, H_{\lambda}^L, \, H_{0}^L . 
\end{align} 
Choose $\beta \in (0,1)$ to be specified later.
Let  $\delta \in (0,   \delta_0^{1/\beta})$. We define the intervals $I = [E-\delta, E+\delta]$ and $\tilde{I}= [E-\delta^\beta, E+\delta^\beta]$.
	Using  the decomposition of  the identity ${\rm{Id}} = 1_{\tilde{I}}(H_0) + 1_{\tilde{I}^c}(H_0)$ we can write 
		\begin{align}  
			{\rm tr}1_I(H_\lambda) &= {\rm tr}1_I(H_\lambda)1_{\tilde{I}}(H_0) +  {\rm tr}1_I(H_\lambda)1_{\tilde{I}^c}(H_0) \label{eq 1}.
		\end{align}
		The first term in  \eqref{eq 1} is bounded  by the simple estimate  
		\begin{align} 
			{\rm tr}1_I(H_\lambda)1_{\tilde{I}}(H_0) \leq \Vert 1_I(H_\lambda) \Vert\; {\rm tr}1_{\tilde{I}}(H_0) \leq {\rm tr}1_{\tilde{I}}(H_0) . \label{first}
		\end{align}
		We decompose the second term in \eqref{eq 1} further
		\begin{align}
			{\rm tr}1_I(H_\lambda)1_{\tilde{I}^c}(H_0) &= {\rm tr}1_I(H_\lambda) (H_\lambda - E)(H_0-E)^{-1}1_{\tilde{I}^c}(H_0) 
\nonumber
\\
			&\qquad- \lambda{\rm tr}1_I(H_\lambda) V(H_0-E)^{-1}1_{\tilde{I}^c}(H_0). \label{eq 2 2}
		\end{align}
	To estimate  the right hand side we will use that by  spectral calculus
		\begin{align}
			&\Vert (H_\lambda-E)1_I(H_\lambda)\Vert \leq \delta\quad  \text{and} \quad \Vert(H_0-E)^{-1}1_{\tilde{I}^c}(H_0)\Vert \leq \frac{1}{\delta^\beta}. \label{spectral calculus}
		\end{align}
		We rewrite  the first term on the right hand side of  \eqref{eq 2 2} in the following way
		\begin{align}   \label{thm:main1:proof:estimate-5}
			& {\rm tr}1_I(H_\lambda) (H_\lambda - E)(H_0-E)^{-1}1_{\tilde{I}^c}(H_0) \\
			&\quad = {\rm tr}1_I(H_\lambda) (H_\lambda - E) |H_0-E|^{-1/2}1_{\tilde{I}^c}(H_0)\frac{|H_0-E|}{H_0-E}1_{\tilde{I}^c}(H_0) |H_0-E|^{-1/2}1_{\tilde{I}^c}(H_0)1_I(H_\lambda) . \nonumber 
		\end{align}
		Define
		\begin{align*}
			A &= 1_I(H_\lambda) (H_\lambda - E) |H_0-E|^{-1/2}1_{\tilde{I}^c}(H_0)\frac{|H_0-E|}{H_0-E}1_{\tilde{I}^c}(H_0) , \\
			B &= |H_0-E|^{-1/2}1_{\tilde{I}^c}(H_0)1_I(H_\lambda) . 
		\end{align*}
		Then we can bound the trace   in  \eqref{thm:main1:proof:estimate-5} by the Hilbert-Schmidt norm, cf. \cite{SR1}, 
		\begin{align*}
			|{\rm tr}1_I(H_\lambda) (H_\lambda - E)(H_0-E)^{-1}1_{\tilde{I}^c}(H_0)| \le \| A\|_2 \|B\|_2 .
		\end{align*}
		Let us first look at the second factor on the right hand side. Here we simply calculate
		\begin{align*}
			\|B\|_2 = \big[ {\rm tr}  1_I(H_\lambda)  1_{\tilde{I}^c}(H_0) |H_0-E|^{-1}1_{\tilde{I}^c}(H_0)1_I(H_\lambda)\big]^{1/2} .
		\end{align*}
		To estimate this  we use 
\begin{align}
 1_{\tilde{I}^c}(H_0) |H_0-E|^{-1}1_{\tilde{I}^c}(H_0) \le \delta^{-\beta}1_{\tilde{I}^c}(H_0) 
\label{thm:main1:proof:estimate-6}
\end{align}
and get by cyclicity
		\begin{align}
			\|B\|_2 \le  \big[ \delta^{-\beta} {\rm tr}  1_I(H_\lambda)  1_{\tilde{I}^c}(H_0)\big]^{1/2} .
		\end{align}
		To estimate the Hilbert-Schmidt norm of  $A$ we first pull out the operator $\frac{|H_0-E|}{H_0-E}1_{\tilde{I}^c}(H_0)$  using $\|ST\|_2 \le \|S\|_2 \|T\|$ with $\left\|\frac{|H_0-E|}{H_0-E}1_{\tilde{I}^c}(H_0)\right\| \le 1$ and obtain 
		\begin{align*}
			\|A\|_2 \le \big[ {\rm tr}   1_{\tilde{I}^c}(H_0) |H_0-E|^{-1/2} 1_I(H_\lambda) (H_\lambda-E)^2  1_I(H_\lambda) |H_0-E|^{-1/2}1_{\tilde{I}^c}(H_0)\big]^{1/2}
		\end{align*}
		Now use \begin{align}
1_I(H_\lambda) (H_\lambda-E)^2  1_I(H_\lambda) \le \delta^2\, 1_I(H_\lambda)
\end{align}
 and cyclicity to estimate
		\begin{align*}
			\|A\|_2 \le \big[ \delta^2 {\rm tr}   1_I(H_\lambda)1_{\tilde{I}^c}(H_0)  |H_0-E|^{-1}1_{\tilde{I}^c}(H_0)1_I(H_\lambda)\big]^{1/2}.
		\end{align*}
		Then  using again \eqref{thm:main1:proof:estimate-6}  we find 
		\begin{align}
			\|A\|_2 \le \big[ \delta^2 \delta^{-\beta} {\rm tr}   1_I(H_\lambda)1_{\tilde{I}^c}(H_0)  \big]^{1/2}.
		\end{align}
		Overall we get 
		\begin{align} \label{secondtermestimate} 
			|{\rm tr}1_I(H_\lambda) (H_\lambda - E)(H_0-E)^{-1}1_{\tilde{I}^c}(H_0)| \le \delta^{1-\beta} {\rm tr}   1_I(H_\lambda)1_{\tilde{I}^c}(H_0) 
		\end{align}
		Note that we can bring this term to the left hand side of \eqref{eq 2 2} since $\delta,\beta \in (0,1)$.
Thus we find 
with the  finite constant $c_{\delta_0,\beta}=(1-\delta_0^{1-\beta})^{-1}$
				\begin{align}
			| {\rm tr}1_I(H_\lambda)1_{\tilde{I}^c}(H_0) | & \leq  c_{\delta_0,\beta} \lambda | {\rm tr}1_I(H_\lambda) V(H_0-E)^{-1}1_{\tilde{I}^c}(H_0) |  .   \label{thm:first:proof:eq3} 
		\end{align}
		Now for the second term in \eqref{eq 2 2} we use \eqref{spectral calculus} and Hölder inequality with $1/p+1/q=1$ for the trace, cf. \cite{simon2005trace},  as well as  for the expectation    
		\begin{align*}
			\mathbf{E}[ \lambda |{\rm tr}1_I(H_\lambda) V(H_0-E)^{-1}1_{\tilde{I}^c}(H_0)|] &\leq  \frac{\lambda}{\delta^\beta}\; \mathbf{E} \bigg(\Vert V\Vert_q \Vert 1_I(H_\lambda)\Vert_p\bigg) \nonumber \\
			&\leq  \frac{\lambda}{\delta^\beta} \bigg(\mathbf{E}\Vert V\Vert_q^q\bigg)^{1/q}  \bigg(\mathbf{E} \Vert 1_I(H_\lambda) \Vert_p^p\bigg)^{1/p}.
		\end{align*}
		Note that 
		\begin{align*}
			\mathbf{E}\Vert V\Vert_q^q = \mathbf{E}{\rm tr}|V|^q = \mathbf{E}\sum_{i\in\Lambda_L}|\omega_i|^q = \sum_{i\in\Lambda_L} \mathbf{E}(|\omega_0|^q) = |\Lambda_L| \mathbf{E}(|\omega_0|^q)
		\end{align*}
		and by $1_I^p = 1_I$   we have
		\begin{align*}
			\Vert1_I(H_\lambda) \Vert_p^p = {\rm{tr}}1_I(H_\lambda).
		\end{align*}
		With that we can use the Wegner estimate \eqref{Wegner estimate} which leads to
		\begin{align}
		&	\lim_{L\to\infty}|\Lambda_L|^{-1}\mathbf{E}[\lambda |{\rm tr}1_I(H_\lambda) V(H_0-E)^{-1}1_{\tilde{I}^c}(H_0)|]\label{thm:first:proof:eq4}  \\
&\leq
			\frac{\lambda}{\delta^\beta} \Vert \omega_0\Vert_q    \bigg(\lim_{L\to\infty}|\Lambda_L|^{-1}\mathbf{E} \, {\rm{tr}}1_I(H_\lambda)\bigg)^{1/p}  \leq \Vert \omega_0\Vert_q (2 \Vert\rho\Vert_\infty)^{1/p} \lambda^{1-1/p} \delta^{1/p-\beta}. \nonumber 
		\end{align}
		with $(\mathbf{E}|\omega_0|^q)^{1/q}=\|\omega_0\|_q$. Since  $\delta \in (0,\delta_0^{1/\beta})$ is small enough,  we can use 
the regular point assumption  \eqref{regularpoint} on  $\tilde I$. Dividing
  \eqref{eq 1}
 by $|\Lambda_L|$ and  taking the limit superior $L\to\infty$
 results  using on the one hand side \eqref{first} together with Remark \ref{remeigenbasis}  and 
on the other \eqref{thm:first:proof:eq3} and  \eqref{thm:first:proof:eq4}  
in
		\begin{align} 
			N_\lambda(I) &\le\Gamma(E,\alpha)\delta^{\alpha\beta} +    c_{\delta_0,\beta}\Vert \omega_0\Vert_q (2 \Vert\rho\Vert_\infty)^{\frac{q-1}{q}} \lambda^{\frac{1}{q}} \delta^{\frac{q-1-q\beta}{q}}.  
		\end{align}
		To calculate $\beta$ we compare this estimate with Wegner's as in \cite{Schenker.2004}. First we look at the region
		\begin{align*}
			\frac{\delta}{\lambda} \ge \lambda^{\frac{1}{q}}\delta^{\frac{q-1-q\beta}{q}} \iff \delta^{\frac{1+q\beta}{q+1}} \ge \lambda
		\end{align*}
		Plugging this into the estimate and comparing the exponents of both $\delta$ terms, to get the same order, results in $\alpha\beta=\alpha/(\alpha+1+\alpha/q)$. One does the same for the other region by switching the inequality and bound the Wegner estimate. This will lead to the same exponent. Overall, by combining both regions, there is a finite constant $ C(E,\rho,q,\alpha)$ such that
		\begin{align*}
			N_\lambda([E-\delta,E+\delta]) \leq C(E,\rho,q,\alpha)\delta^{\frac{\alpha}{\alpha+1+\alpha/q}}.
		\end{align*}
		for all $0<\lambda$ and $\delta\in[0, \delta_0^{\alpha+1+\alpha/q})$. To extend this to all $\delta\ge0 $, we use the finiteness of the measure $N_\lambda$ and get for $\delta^{\frac{1}{\alpha+1+\alpha/q}}\ge\delta_0$
		\begin{align*}
			N_\lambda([E-\delta,E+\delta]) \leq 1\le \delta_0^{-\alpha} \delta^{\frac{\alpha}{\alpha+1+\alpha/q}}.
		\end{align*}
		Thus the result for general  $\lambda, \delta  \geq 0$ now  follows by combining the 
different cases.
	\end{proof}
	\begin{remark}
		\normalfont  The bound \eqref{ineqforboundfirst}  is the same as in \cite{Schenker.2004}, except for a different constant and that it holds for $q \geq 1$ rather than $q \geq 2$.
	\end{remark}

	\section{Lower Moment Estimates}
\label{sec:lowermoment} 
	We show in this chapter some improved Hölder estimates. These are based on a better estimation of trace terms of the form 
	\begin{align*}
		\mathbf{E}\,{\rm{tr}}[ F_1(H_0^L)V_L F_2(H_0^L)V_LF_3(H_0^L) \cdots V_L F_{n+1}(H_0^L)]
	\end{align*}
	for some bounded measurable functions $F_i$. 
For simplicity we restrict our analysis to  regular points. 
 We will  show  in Appendix \ref{sec:integralbounds}  the following estimates for  a regular point $E$: 
For all $\varepsilon \in(0, \delta_0]$ we have
	\begin{align*}
		&\int_{\{|\mathfrak{e}(s)-E|\ge \varepsilon\}}(\mathfrak{e}(s)-E)^{-2}\mathrm{d}\widehat{s} \leq  C_{2,E} \,\varepsilon^{-1},  \\
		&\int_{\{|\mathfrak{e}(s)-E|\ge \varepsilon\}}|\mathfrak{e}(s)-E|^{-1}\mathrm{d}\widehat{s} \leq  C_{1,E} \,(1+|\ln \varepsilon|),  \\
		&\int_{\T^d}1_{[E-\varepsilon, E+\varepsilon]}(\mathfrak{e}(s))\mathrm{d}\widehat{s} \leq C_E\, \varepsilon.
	\end{align*}
These bounds will be used to improve  the estimates on the  resolvents.
	We will start with random variables for which we only require  $\mathbf{E}[\omega_0^2]<\infty$.
	\begin{theorem} \label{Improved1}
		Assume that $ \mathbf{E}[\omega_0^2]<\infty$. Let $E$ be a regular point.
		Then we have
		\begin{align}
			N_\lambda([E-\delta,E+\delta])\leq C(\rho, E)\delta^{\frac{1}{2}}
		\end{align}
		for all $\lambda,\delta\ge 0$.
	\end{theorem}
	\begin{proof}
		The proof follows the same steps as that of Theorem 
\ref{regpoint alpha} except that  we estimate the expectation of the second term of the right hand side  \eqref{eq 2 2}
as follows. Again we suppress the $L$ label and use the notation \eqref{shorhandnot}.
Using Cauchy-Schwarz inequality for the trace  and the expectation ${\bf E}$ we find 
		\begin{align*}
			&\lambda \mathbf{E} |{\rm tr}1_I(H_\lambda) V(H_0-E)^{-1}1_{\tilde{I}^c}(H_0)| \\
			&\quad\le \lambda (\mathbf{E}{\rm tr} 1_I(H_\lambda))^{1/2}  \cdot (\mathbf{E}{\rm tr} 1_{\tilde{I}^c}(H_0) (H_0-E)^{-1}V^2(H_0-E)^{-1}1_{\tilde{I}^c}(H_0) )^{1/2} .
		\end{align*}
		For the second factor we use $\mathbf{E} [V^2]= \mathbf{E}[\omega_0^2]{\rm Id}$. Dividing by the volume and taking the limit superior results in
		\begin{align*}
			N_\lambda(I) \le \Gamma(E) \delta^\beta +C \delta^{1/2} \lambda^{1/2} \left(  \limsup_{L\to\infty} |\Lambda_L|^{-1} {\rm tr} (H_0-E)^{-2}1_{\tilde{I}^c}(H_0)\right)^{1/2} .
		\end{align*}
Using the definition of the measure $N_0^L$ \eqref{finitedospushforwards}, we can write the term in the bracket as
		\begin{align*}
			 \limsup_{L\to\infty} |\Lambda_L|^{-1} {\rm tr} (H_0-E)^{-2}1_{\tilde{I}^c}(H_0) = \limsup_{L\to\infty} \int_{|x-E|>\delta^\beta} (x-E)^{-2}dN_0^L(x).
		\end{align*}
		We bound the integrand above by $1_{\{|x-E|\ge \delta^\beta\}}(x-E)^{-2}$. This function is bounded and upper semicontinuous, thus we can apply the Portmanteau theorem, cf.   \eqref{uppersemig}, in combination with the bound \eqref{Int 1} to obtain
		\begin{align*}
			\limsup_{L\to\infty} \int_{|x-E|>\delta^\beta} (x-E)^{-2}dN_0^L(x) \le \int_{|x-E|\ge \delta^\beta} (x-E)^{-2}dN_0(x)\le C_E\delta^{-\beta}.
		\end{align*} 
		 Inserting results in
		\begin{align*}
			N_\lambda(I) \le \Gamma(E) \delta^\beta +\widetilde C \delta^{(1-\beta)/2} \lambda^{1/2}  .
		\end{align*}
		Now as in the proof of  Theorem \ref{regpoint alpha} we perform  the comparison   with the Wegner estimate and obtain  $\beta = \frac{1}{2}$. Thus there exists a finite constant $ C(\rho, E) = \max\{\Gamma(E)+\widetilde C, 2\|\rho\|_\infty\}$ such that
		\begin{align}
			N_\lambda(I) \le  C(\rho, E) \delta^{\frac{1}{2}} 
		\end{align}
		for all $\lambda>0$ and $\delta \in[0, \delta_0^{2})$. For $\delta \ge \delta_0^{2}$ we do the same argument as in Theorem \ref{regpoint alpha}. The statement now follows by combining the $\lambda=0$ and $\lambda>0$ cases.
	\end{proof}
	\begin{remark}\normalfont
		We can extend this result to regular points of order $\alpha$. For this we use the estimate \eqref{Int a 1}, which implies $$\int_{\{|\mathfrak{e}(s)-E|\ge \varepsilon\}}(\mathfrak{e}(s)-E)^{-2}\mathrm{d}\widehat{s} \leq  C_{E,\alpha} \,\varepsilon^{\alpha-2}.$$
			Following the same steps as in the proof of  Theorem \ref{Improved1} we get 
			\begin{align*}
				N_\lambda(I) \le \Gamma(E,\alpha)\delta^{\alpha \beta} +C_{E,\alpha,\beta}\lambda^{1/2}\delta^{1/2 +\alpha\beta/2-\beta }
			\end{align*} 
			for all $\lambda>0$ and $\delta\in(0,\delta^{1/\beta})$.
			Again we calculate $\beta$ by comparing with Wegner's estimate and obtain, by combining all regions,
			\begin{align*}
				N_\lambda([E-\delta,E+\delta])\le C(\rho,E,\alpha)\delta^{\frac{\alpha}{\alpha+1}}
			\end{align*}
			for all $\lambda,\delta\ge 0$.
	\end{remark}
	We can get a better H\"older exponent by assuming the finiteness of  higher moments.
	\begin{theorem}\label{Improved}
		Assume that 
 $\mathbf{E}[\omega_0]=0$ and  $\mathbf{E}[\omega_0^4]<\infty$. Let $E$ be a regular point.
		Then 
		\begin{align}
			N_\lambda([E-\delta,E+\delta])\leq C(\rho, E)\delta^{\frac{4}{7}}
		\end{align}
		for all $\lambda,\delta\ge0$.
	\end{theorem}

	\begin{proof} If $\lambda = 0$, then the bound follows immediately from the definition. Thus we can assume $\lambda > 0$. The case $\delta =0$ follows by Wegner's estimate  \eqref{Wegner estimate}, hence let $\delta > 0$. 
Again we adapt the notation   \eqref{shorhandnot}.  By the regular point assumption there 
exists a $\delta_0 \in (0,1)$ such that \eqref{regularpoint1}  holds. Let $\beta \in (0,1)$, which well be determined 
later.  Let $\delta \in (0,\delta_0^{1/\beta})$. We write $ I=[E-\delta,E+\delta]$ and $\tilde I = [E- \varepsilon,E+ \varepsilon]$ for $\varepsilon=\delta^\beta$, as well as\begin{align} \label{defofG1} G_1(H_0) = (H_0-E)^{-1}1_{\tilde{I}^c}(H_0) \end{align}  for notational compactness.
		We decompose as follows
		\begin{align} 
			{\rm tr}1_I(H_\lambda)
			&={\rm tr}1_I(H_\lambda)1_{\tilde I}(H_0) +{\rm tr}1_I(H_\lambda) (H_\lambda - E)G_1(H_0) \nonumber\\ 
			&\qquad - \lambda {\rm tr}1_I(H_\lambda) 1_{\tilde I}(H_0) V G_1(H_0) - \lambda {\rm tr}1_I(H_\lambda) 1_{\tilde I^c}(H_0) V G_1(H_0)\nonumber \\
			&={\rm tr}1_I(H_\lambda)1_{\tilde I}(H_0) +{\rm tr}1_I(H_\lambda) (H_\lambda - E)G_1(H_0) \label{1}\\
			&\qquad- \lambda {\rm tr}1_I(H_\lambda)1_{\tilde I}(H_0) V G_1(H_0) \label{2} \\
			&\qquad -\lambda {\rm tr}1_I(H_\lambda)(H_\lambda-E)G_1(H_0)VG_1(H_0)\label{3} \\
			&\qquad +\lambda^2 {\rm tr}1_I(H_\lambda)VG_1(H_0) VG_1(H_0). \label{4}
		\end{align}
		The  first term  in line \eqref{1}   can be bounded, as in the proof of Theorem \ref{regpoint alpha}, by  \eqref{first}. 
The second term in line \eqref{1} can be bounded   by \eqref{secondtermestimate}  and thus be taken to the left hand side, since 
 $\delta,\beta \in (0,1)$, again, as in the proof of Theorem \ref{regpoint alpha}.
		
	To estimate  \eqref{2} we use Cauchy-Schwarz inequality for the trace and the expectation  
		\begin{align}
			\limsup_{L\to\infty}\frac{\mathbf{E}\eqref{2}}{|\Lambda_L|} &\leq \limsup_{L\to\infty}\frac{\lambda}{|\Lambda_L|} \mathbf{E}(\Vert 1_I(H_\lambda)\Vert_2 \cdot \Vert 1_{\tilde I}(H_0) V G_1(H_0) \Vert_2) \nonumber \\
			&\leq \limsup_{L\to\infty}\frac{\lambda}{|\Lambda_L|} \left( \mathbf{E}{\rm tr}1_I(H_\lambda) \right)^{\frac{1}{2}} \cdot \left(  \mathbf{E} {\rm tr} |1_{\tilde I}(H_0) V G_1(H_0)|^{2} \right)^{\frac{1}{2}} \nonumber \\
			&\leq C_{\rho,E}\delta^{\frac{1}{2}}\lambda^{1-\frac{1}{2}} ,
		\end{align}
		where in the last line we used Wegner's estimate (Theorem  \ref{thm:wegner}) for the first factor and  
  \eqref{Int 3} together with  \eqref{Int 1} for the second factor. 
		For \eqref{3} we find  using Cauchy-Schwarz as before
		\begin{align}
		\limsup_{L\to\infty}\frac{\mathbf{E}\eqref{3}}{|\Lambda_L|} &\leq \limsup_{L\to\infty}\frac{\delta\lambda}{|\Lambda_L|} \mathbf{E}(\Vert 1_I(H_\lambda)\Vert_2 \cdot \Vert G_1(H_0) V G_1(H_0) \Vert_{2}) \nonumber \\
			&\leq \limsup_{L\to\infty}\frac{\delta\lambda}{|\Lambda_L|} \left( \mathbf{E}{\rm tr}1_I(H_\lambda) \right)^{\frac{1}{2}} \cdot \left(  \mathbf{E} {\rm tr} |G_1(H_0) V G_1(H_0)|^{2} \right)^{\frac{1}{2}} \nonumber \\
			&\leq  C_{\rho,E}\delta^{1+\frac{1}{2}}\lambda^{1-\frac{1}{2}} \varepsilon^{-1} , 
		\end{align}
where in the last line we used Wegner's estimate and  \eqref{Int 1}.
		Now we  use  a similar estimate for the last bound \eqref{4} in combination with Wegner's estimate,  \eqref{Int 1}, and \eqref{Int 2}
		\begin{align}
			\limsup_{L\to\infty}\frac{\mathbf{E}\eqref{4}}{|\Lambda_L|} &\leq \limsup_{L\to\infty}\frac{\lambda^2}{|\Lambda_L|} \mathbf{E}(\Vert 1_I(H_\lambda)\Vert_2  \cdot \Vert VG_1(H_0) V G_1(H_0) \Vert_{2}) \nonumber \\
			&\leq \limsup_{L\to\infty}\frac{\lambda^2}{|\Lambda_L|} \left( \mathbf{E}{\rm tr}1_I(H_\lambda) \right)^{\frac{1}{2}} \cdot \left(  \mathbf{E} {\rm tr} G_1(H_0) V G_1(H_0)V^2G_1(H_0) V G_1(H_0)\right)^{\frac{1}{2}} \nonumber \\
			&\leq C_{\rho,E} \delta^{\frac{1}{2}} \lambda^{\frac{3}{2}} \varepsilon^{-1},
		\end{align}
		where in the last line we calculated the trace and used the worst case bound $\varepsilon^{-2}$ from all pairings, see Lemma \ref{boundonfourV}.  
 So collecting the above  estimates we arrive at  the following inequality
		\begin{align}
			N_\lambda(I) &\leq \Gamma(E)\varepsilon + C_1 \delta^{\frac{1}{2}}\lambda^{1-\frac{1}{2}}  
			+  C_2\delta^{1+\frac{1}{2}}\lambda^{1-\frac{1}{2}}  \varepsilon^{-1} + C_3   \delta^{\frac{1}{2}} \lambda^{\frac{3}{2}} \varepsilon^{-1}.
		\end{align}
		Note that $\delta/\varepsilon < 1$ and so we can absorb the third into the second term. Since $\varepsilon = \delta^\beta$, we can write
		\begin{align}
			N_\lambda(I) &\le \Gamma(E)\delta^\beta +(C_1+C_2)\delta^{\frac{1}{2}}\lambda^{\frac{1}{2}}  + C_3 \delta^{\frac{1-2\beta}{2}}\lambda^{\frac{3}{2}}.
		\end{align}
		We  consider  again  two complementary parameter regions. The first is
		\begin{align} \label{firstregion-3} 
			\frac{\delta}{\lambda} \ge\delta^{\frac{1-2\beta}{2}}\lambda^{\frac{3}{2}} \iff \delta^{\frac{1+2\beta}{5}} \ge \lambda.
		\end{align}
		Inserting this into the IDS bound we get
		\begin{align}
			N_\lambda(I) &\le \Gamma(E)\delta^\beta + (C_1+C_2)\delta^{\frac{3+\beta}{5}} +C_3\delta^{\frac{4-2\beta}{5}} \nn \\
			&\le \Gamma(E)\delta^\beta  +(C_1+C_2+C_3)\delta^{\frac{4-2\beta}{5}},
		\end{align}
		since we want at least $1/2 \le \beta$. 
		By comparing the exponents we need
		\begin{align*}
			\beta = \frac{4}{7}
		\end{align*}
		to get the same order.
		
		Now we consider the  region complemtary to   \eqref{firstregion-3}, that is   the reverse inequality 
		\begin{align*}
			\delta^{\frac{1+2\beta}{5}} \le \lambda
		\end{align*}
		and plug it into  Wegner's estimate with $\beta=4/7$ and get
		\begin{align*}
			N_\lambda(I)\le 2\|\rho\|_\infty \delta^{\frac{4}{7}}.
		\end{align*}
		
		By combining the two regions, we see that there exists a finite constant $C(\rho,E) = \max\{\Gamma(E)+C_1+C_2+C_3, 2\|\rho\|_\infty\}$, such that
		\begin{align*}
			N_\lambda([E-\delta,E+\delta]) \le C(\rho,E) \delta^{\frac{4}{7}}
		\end{align*}
	for all $\lambda > 0$ and $\delta\in[0, \delta_0^{7/4})$. For large $\delta$ the desired estimate now follows by the same argument as in the proof of Theorem \ref{regpoint alpha}. Thus the statement holds by combining the $\lambda=0$ and $\lambda>0$ cases  with  suitable constants.
	\end{proof}

 \begin{lemma}  \label{boundonfourV} Let the assumptions be as in Theorem \ref{Improved}, and let $E$ be a regular point such that 
 \eqref{regularpoint1}  holds for $\delta_0 \in (0,1)$.
Then there  exists a constant $C$ such that for all $\varepsilon  \in (0,\delta_0]$ 
\begin{align*} 
\limsup_{L \to \infty} |\Lambda_L|^{-1}  \mathbf{E} {\rm tr} G_1(H^L_0) V_L G_1(H^L_0)V_L^2G_1(H_0^L) V_L G_1(H_0^L)  \leq C  \varepsilon^{-2}   .
\end{align*} 
\end{lemma} 
\begin{proof} We supress the $L$-labels and adapt the convention  \eqref{shorhandnot}. 
Since ${\bf E}[\omega_0 ] = 0$ we find   calculating the expectation 
\begin{align*}
& \mathbf{E} {\rm tr} G_1(H_0) V G_1(H_0)V^2G_1(H_0) V G_1(H_0) \\
&  = \left(  {\bf E}[\omega_0^4] -  ({\bf E}[\omega_0^2])^2\right)  I_1 +  [{\bf E}(\omega_0^2)]^2  I_2 , 
\end{align*}  
where  
\begin{align*}
I_1  & := |\Lambda_L|^{-1}\sum_x  {\rm tr} [G_1(H_0)  P_x G_1(H_0) P_x G_1(H_0) P_x  G_1(H_0) ]  \\
  I_2   & := 
|\Lambda_L|^{-1}\sum_{x,y}   {\rm tr} [ G_1(H_0) P_x G_1(H_0) P_y G_1(H_0)  P_x  G_1(H_0) ]
\end{align*} 
with   $P_x =\delta_x  \langle \delta_x , \cdot \rangle $ denoting the orthogonal projection onto $\delta_x$. 
We calculate using translation invariance 
\begin{align*}
I_1  & = |\Lambda_L|^{-1}\sum_x  {\rm tr} [ G_1(H_0)  P_x G_1(H_0) P_x G_1(H_0) P_x  G_1(H_0) ] \\
& =|\Lambda_L|^{-3} ( {\rm tr} [G_1(H_0) ])^2 {\rm tr} [ G_1(H_0)^2] ,
\end{align*} 
\begin{align*} 
  I_2 
& = |\Lambda_L|^{-1}\sum_{x}   {\rm tr} G_1(H_0) \langle \delta_x  ,  G_1(H_0)  G_1(H_0)  \delta_x  \rangle P_x    G_1(H_0) \\
& = \left( |\Lambda_L|^{-1}  {\rm tr} [ G_1(H_0)^2  ] \right)^2  .
\end{align*} 
Now we estimate the  limit superior by  means of monotonicity, \eqref{uppersemig},  \eqref{Int 1}, and   \eqref{Int 2} 
\begin{align*}
\limsup_L I_1  & \leq \limsup_L  \left(  |\Lambda_L|^{-1}{\rm tr}  [ |H_0-E|^{-1} 1_{\overline{\tilde{I}^c}}(H_0)  ) ]\right)^2 \left(
 |\Lambda_L|^{-1} {\rm tr} [ |H_0-E|^{-2} 1_{\overline{\tilde{I}^c}}(H_0) ] \right) \\
& \leq C  ( 1  + | \ln \varepsilon | )^{2} \varepsilon^{-1}  ,
\end{align*} 
\begin{align*} 
\limsup_L  I_2 
& \leq  \limsup_L \left( |\Lambda_L|^{-1}  {\rm tr} [  |H_0-E|^{-2} 1_{\overline{\tilde{I}^c}}(H_0)  ] \right)^2   \leq C  \varepsilon^{-2}  .
\end{align*} 
Inserting estimates yields the claim. 
\end{proof} 

	\section{Higher  Moment Estimates}
\label{sec:HigherMomentEstimates}
	In this section  we extend the  methods of the previous sections to expansions of arbitrary high order. We first proof the following preliminary result. 
	\begin{lemma}
		Let  $\mathbf{E}[\omega_0] = 0  $, and let  $n \in \N$ with  $n \geq 3$ be such that    $ \mathbf{E}[\omega_0^{2n}] < \infty$.
Suppose $E$ is a regular point and let $\delta_0 \in (0,1) $  and $\Gamma(E)$ be  such that  \eqref{regularpoint1} holds for all $\delta \in [0,\delta_0]$.
Then for  any $\beta \in (0,1)$ there are constants $C_k(\beta)$ such that  
		\begin{align}
			N_\lambda([E-\delta,E+\delta]) &\le \Gamma(E) \delta^\beta + C_1(\beta)\lambda^{1/2}\delta^{1/2} + \sum_{k=2}^{n-1}C_k(\beta) \lambda^{\frac{2k-1}{2}} \delta^{\frac{1-(k-1)\beta}{2}}(1+\beta|\ln\delta|)^{k} \nn \\
			&\qquad +C_n(\beta) \lambda^{\frac{2n-1}{2}} \delta^{\frac{1-n\beta}{2}}(1+\beta|\ln\delta|)^{n}
		\end{align}
for all $\lambda>0$ and $\delta \in (0, \delta_0^{1/\beta}]$.
	\end{lemma}
	\begin{proof}
		The proof  follows  the ideas of 
the proof of 
Theorem \ref{regpoint alpha}, but  we expand further up to the $n$'th order. 
We supress the $L$-labels and adapt   \eqref{shorhandnot}. 
Let $I=[E-\delta,E+\delta]$ and $\tilde{I}=[E-\varepsilon,E+\varepsilon]$ for $\varepsilon>\delta$. 
If follows from \eqref{eq 1}, \eqref{first}, and \eqref{secondtermestimate}  that 
		\begin{align} \label{higherordereststart} 
			\mathrm{tr}1_I(H_\lambda)\le \mathrm{tr}1_{\tilde{I}}(H_0)+\frac{\varepsilon}{\varepsilon-\delta}\lambda |\mathrm{tr}1_I(H_\lambda)VG_1(H_0)|  ,
		\end{align}
where we used the notation introduced in  \eqref{defofG1}.
		Define 
		\begin{align*}
			R_k&=\mathrm{tr}1_I(H_\lambda)[VG_1(H_0)]^k,\\
			A_k&=\mathrm{tr}1_I(H_\lambda)1_{\tilde{I}}(H_0)[VG_1(H_0)]^k,\\
			B_k&=\mathrm{tr}1_I(H_\lambda)(H_\lambda-E)G_1(H_0)[VG_1(H_0)]^k.
		\end{align*}
		Inserting $ {\rm Id} = ( H_\lambda - E - \lambda V ) G_1(H_0)   + 1_{\tilde{I}}(H_0)$ into $R_k$, we obtain the relation  
		\begin{align*}
			R_k =A_k +B_k-\lambda R_{k+1}.
		\end{align*}
		Iterating this identity from $k=1$ to $n-1$, we find 
		\begin{align*}
			R_1=\sum_{k=1}^{n-1}(-\lambda)^{k-1}(A_k+B_k)+(-\lambda)^{n-1}R_n.
		\end{align*}
		Inserting this into \eqref{higherordereststart}  we can bound
		\begin{align*}
				\mathrm{tr}1_I(H_\lambda)\le \mathrm{tr}1_{\tilde{I}}(H_0)+\frac{\varepsilon}{\varepsilon-\delta}\sum_{k=1}^{n-1}\lambda^k(|A_k|+|B_k|)+\frac{\varepsilon}{\varepsilon-\delta} \lambda^n|R_n|.
		\end{align*}
		We divide by the volume, apply the expectation and take the limit superior. For $A_k$ and $B_k$ we use Cauchy-Schwarz inequality and Corollary \ref{Contractionapp 2}. 
Hence
		\begin{align*}
			\limsup_{L\to\infty} |\Lambda_L|^{-1}\mathbf{E}|A_k|&\le C_k (N_\lambda([E-\delta,E+\delta]))^{1/2}\varepsilon^{-\frac{k-1}{2}}(1+|\ln\varepsilon|)^k, \\
			\limsup_{L\to\infty} |\Lambda_L|^{-1}\mathbf{E}|B_k|&\le C_k (N_\lambda([E-\delta,E+\delta]))^{1/2} \,\frac{\delta}{\varepsilon}\,\varepsilon^{-\frac{k-1}{2}}(1+|\ln\varepsilon|)^k.
		\end{align*}
        We can absorb the second estimate, since $\delta/\varepsilon <1$. For $k=1$ we do not need the logarithmic correction, cf. 
 Corollary \ref{Contractionapp 2}.  For the remainder $R_k$ we use Cauchy-Schwarz inequality and Corollary \ref{Contractionapp 1}, and get
        \begin{align*}
        		\limsup_{L\to\infty} |\Lambda_L|^{-1}\mathbf{E}|R_n|\le C_n (N_\lambda([E-\delta,E+\delta]))^{1/2}\varepsilon^{-\frac{n}{2}}(1+|\ln\varepsilon|)^n.
        \end{align*}
        Overall we arrive  using Wegner's estimate  at
        \begin{align*}
        	N_\lambda(I)&\le \Gamma(E)\varepsilon+ 2\frac{\varepsilon}{\varepsilon-\delta}C_1 \lambda^{1/2}\delta^{1/2}+2\frac{\varepsilon}{\varepsilon-\delta}\sum_{k=2}^{n-1}C_k \lambda^{k-1/2}\delta^{1/2}\varepsilon^{-\frac{k-1}{2}}(1+|\ln\varepsilon|)^k \\
        	&\qquad +\frac{\varepsilon}{\varepsilon-\delta}C_n \lambda^{n-1/2}\delta^{1/2}\varepsilon^{-n/2}(1+|\ln\varepsilon|)^n.
        \end{align*}
         Now using  $\varepsilon =\delta^\beta \le \delta_0$ the desired estimate follows. 
	\end{proof}
	With this expansion given, we now want to improve the Hölder exponent for large $n$. 
	\begin{theorem} \label{thm:genorder} 
		Let $n \in \N$ with $n\ge3$. Assume that  $\mathbf{E}[\omega_0]=0$ and  $\mathbf{E}[\omega_0^{2n}]<\infty$. Let $E$ be a regular point. Then there exists a finite constant $C(E,\rho, n)$ such that 
		\begin{align}
			N_\lambda([E-\delta,E+\delta]) \le C(E,\rho, n) \delta^{\frac{2n}{3n+1}}\left(1+\frac{2n}{3n+1} |\ln \delta|\right)^{\frac{2n}{2n+1}}
		\end{align}
		for all $\lambda,\delta\ge0$.
	\end{theorem}
	\begin{proof}
		The case $\lambda=0$ follows by definition. Hence we consider $\lambda>0$ and $\delta\in[0,\delta_0^{1/\beta})$ as in Theorem \ref{regpoint alpha}. For $\delta=0$ this follows by Wegner's estimate \ref{Wegner estimate}, thus let $\delta>0$. We start by using the previous lemma to bound
		\begin{align*}
			N_\lambda(I) &\le \Gamma(E) \delta^\beta + C_1\lambda^{1/2}\delta^{1/2} + \sum_{k=2}^{n-1}C_k \lambda^{\frac{2k-1}{2}} \delta^{\frac{1-(k-1)\beta}{2}}(1+\beta|\ln\delta|)^{k} \nn \\
			&\qquad +C_n \lambda^{\frac{2n-1}{2}} \delta^{\frac{1-n\beta}{2}}(1+\beta|\ln\delta|)^{n}
		\end{align*}
		with $\beta(n) =2n/(3n+1)$. We first look at the region
		\begin{align*}
			\frac{\delta}{\lambda}\ge\lambda^{\frac{2n-1}{2}} \delta^{\frac{1-n\beta}{2}}(1+\beta|\ln\delta|)^{n} \iff \delta^{\frac{1+n\beta}{2n+1}}(1+\beta|\ln\delta|)^{-\frac{2n}{2n+1}} \ge \lambda
		\end{align*}
		In that region we can bound $\lambda$ from above by the above inequality and find 
		\begin{align*}
			N_\lambda(I) &\le \Gamma(E) \delta^{\beta} +C_1\delta^{\frac{2(n+1)+n\beta}{2(2n+1)}}(1+\beta|\ln\delta|)^{-\frac{n}{2n+1}}+ \sum_{k=2}^{n-1}C_k \delta^{\frac{2(n+k)+\beta(n-k+1)}{2(2n+1)}}(1+\beta|\ln\delta|)^{\frac{n+k}{2n+1}}\\
			&\qquad +C_n \delta^{\frac{4n-2n\beta }{2(2n+1)}}(1+\beta|\ln\delta|)^{\frac{2n}{2n+1}} \\
			&\le \bigg[ \Gamma(E) \delta^{\beta} +C_1\delta^{\frac{2(n+1)+n\beta}{2(2n+1)}}+ \sum_{k=2}^{n-1}C_k \delta^{\frac{2(n+k)+\beta(n-k+1)}{2(2n+1)}}
			+C_n \delta^{\frac{4n-2n\beta }{2(2n+1)}}\bigg](1+\beta|\ln\delta|)^{\frac{2n}{2n+1}} \\
			&\le ( \Gamma(E)+\widetilde C_n )\delta^{\frac{2n }{3n+1}}\left(1+\frac{2n}{3n+1} |\ln \delta|\right)^{\frac{2n}{2n+1}} ,
		\end{align*}
		where we used $\beta=\frac{2n}{3n+1}$ and $\widetilde C_n =n \max_{i=1,\dots,n} C_i$.  Note that the choice of $\beta(n)$ balances the exponent in the first and last term. Furthermore, we where able to bound all other terms above by the last one. For the other region we bound the Wegner estimate and get the same orders up to some constant. Overall there exists a constant $C(E,\rho,n) = \max\{ \Gamma(E)+\widetilde C_n, 2\|\rho\|_\infty\}$, such that
		\begin{align*}
			N_\lambda(I) \le C(E,\rho,n) \delta^{\frac{2n}{3n+1}}\left(1+\frac{2n}{3n+1} |\ln \delta|\right)^{\frac{2n}{2n+1}}
		\end{align*}
		for all $\lambda >0$ and $\delta\in [0, \delta_0^{\frac{3n+1}{2n}})$. For large $\delta$ the desired bound  follows by 
an  argument analogous to the one given in the proof of  Theorem \ref{regpoint alpha}.
	\end{proof}

From the theorem we immediatly obtain the following corollary. 
	\begin{corollary} \label{corollarymain} 
		Assume that $\mathbf{E} [\omega_0] =0$ and $\mathbf{E}[|\omega_0|^n ] < \infty$  for all $n\in \N$. Let $E$ be a regular point of $H_0$. Then for every $0<\beta<2/3$ there exists a constant $C_{E,\beta}<\infty$, such that
		\begin{align*}
			N_\lambda([E-\delta,E+\delta])\le C_{E,\beta}\delta^\beta
		\end{align*}
		for all $\lambda,\delta\ge0$.
	\end{corollary}
	\section{Bounding Expansion Terms and Feynman Graphs}

\label{sec:feynmanbounds} 
	In this section we  show  bounds  for  averaged traces of the form  
	\begin{align}\label{trace}
		\mathbf{E}\,{\rm{tr}}[ F_1(H_0^L)V_L F_2(H_0^L)V_LF_3(H_0^L) \cdots V F_{n+1}(H_0^L)].
	\end{align}

	\begin{theorem} \label{main:graphbound} 
		Let $n \in \N$.  Assume   $M_n := \mathbf{E}[|\omega_0|^n]<\infty$, and  assume $\{F_i\}_{i=1,\dots,n+1}$ are bounded measurable functions. Let $\mathcal{A}_n$ be the set of all partitions of $\{1,\dots,n\}$, where we define for each $A\in \mathcal{A}_n$ the set $J_A=\{\max a : a\in A\}$ and $I_A=\{1,\dots,n\}\setminus J_A$. Then 
		\begin{align*}
			&\big|\mathbf{E}\,{\rm{tr}}[ F_1(H_0^L)V_L F_2(H_0^L)V_LF_3(H_0^L) \cdots V_L F_{n+1}(H_0^L)] \big| \\
			&\le \sum_{q\in \hat{\Lambda}_L}|F_1(\mathfrak{e}_L(q))F_{n+1}(\mathfrak{e}_L(q))| \sum_{A\in \mathcal{A}_n} |c_A| \prod_{j\in J_A\setminus\{n\}} \| F_{j+1}\|_\infty \, \prod_{j\in I_A}\left[|\Lambda_L|^{-1}\sum_{k\in \hat{\Lambda}_L}|F_{j+1}(\mathfrak{e}_L(k))| \right] , 
		\end{align*}
		where  $c_A$  are constants depending on the product of moments $ \prod_{a \in A}  M_{|a|}$.
	\end{theorem}
	\begin{proof}
		To prove the theorem we work  in momentum space.
		That is, we expand with respect 
		to the eigenbasis introduced in \eqref{eigenbasisfinite} 
		\begin{align} \label{eigbasis-2} 
			\varphi_k(x) = |\Lambda_L|^{-1/2} e^{  i k \cdot x} , \quad x \in \Lambda_L ,
		\end{align}
		with $k \in \widehat{\Lambda}_L$. 
		We obtain 
		\begin{align*}
			\langle \varphi_k , V_L \varphi_{p} \rangle = |\Lambda_L|^{-1} \sum_{x  \in \Lambda_L} \omega_x e^{  i (p-k) \cdot x } .
		\end{align*} 
		Let us define 
		\begin{align*}
			\hat{V}(k) := |\Lambda_L|^{-1} \sum_{x \in \Lambda_L} \omega_x e^{  i k \cdot x } .
		\end{align*} 
		Then 
		\begin{align*}
			{\bf E}  ( \hat{V}(k) \hat{V}(p) )  & ={\bf E} ( |\Lambda_L|^{-2} \sum_{x \in \Lambda_L} \omega_x  e^{  i  k \cdot x}
			\sum_{x' \in \Lambda_L} \omega_{x'}  e^{  i  p  \cdot x' } )  \\
			&=   |\Lambda_L|^{-1}\mathrm{Var}(\omega_0)  \delta_{k,-p} +(\mathbf{E}\omega_0)^2\delta_{k,0}\delta_{p,0}.
		\end{align*} 
		Note that
		$\delta_{q,p} = \delta_{q-p,0}
		$.
		To calculate the expectation for the product of more than two terms, let $ \mathcal{A}_n$ be the set of all partitions of $\{1,\dots,n\}$. We define for $A \in \mathcal{A}_n$ the two functions $\chi_A$ and $\tilde{\chi}_A$ on $(\mathbb{Z}^d)^n$
		as follows. Let $(z_1,...,z_n) \in (\Z^d)^n$.  First,  we set  
		\begin{align} 
			\label{X1}
			\chi_A(z_1,....,z_n) := \prod_{a \in A} \bigg[ \prod_{(j,l) \in a \times a } 1_{\{  z_j = z_l \}} \bigg].
		\end{align} 
		Second,  we define  
		\begin{align} 
			\label{X2}
			\tilde{\chi}_A(z_1,....,z_n) := \chi_A(z_1,...,z_n)  \prod_{\substack{ ( a, b )   \in A \times A :  \\ a \neq b}}  \bigg[ \prod_{(j,l) \in a \times b} 1_{\{  z_j \neq  z_l \}} \bigg]   ,
		\end{align} 
		Observe that for any $(z_1,...,z_n) \in (\Z^d)^n$ we have 
		\begin{align}
			\label{upartition}
			\sum_{A \in \mathcal{A}_n}  \tilde{\chi}_A(z_1,...,z_n)  = 1 .
		\end{align}
		With this small preamble, we can express the expectation value  term wise and show the following lemma.
		Using that $1_{\{  z_j \neq  z_l \}} =  1 - 1_{\{  z_j =   z_l \}} $ and 
		the binomial theorem we find 
		\begin{align*}
			& \tilde{\chi}_A(z_1,....,z_n) \\
			& = \chi_A(z_1,...,z_n)  \prod_{\substack{ ( a, b )   \in A \times A :  \\ a \neq b}}  \bigg[ \prod_{(j,l) \in a \times b} (1 - 1_{\{  z_j =   z_l \}} ) \bigg]   \\
			& = \chi_A(z_1,...,z_n)  \prod_{\substack{ ( a, b )   \in A \times A :  \\ a \neq b}}  \bigg[ 1 + \sum_{k=1}^{|a| |b|} \binom{ |a||b| }{ k} (- 1)^k 1_{\{  z_j =   z_l : j \in a, l \in b \}}  \bigg]   \\
			& = \chi_A(z_1,...,z_n)  \prod_{\substack{ ( a, b )   \in A \times A :  \\ a \neq b}}  \bigg[ 1 -  1_{\{  z_j =   z_l \}} + \sum_{k=0}^{|a| |b|} \binom{ |a||b| }{ k} (- 1)^k 1_{\{  z_j =   z_l : j \in a, l \in b\}}  \bigg]   \\
			& = \chi_A(z_1,...,z_n)  \prod_{\substack{ ( a, b )   \in A \times A :  \\ a \neq b}}  \bigg[ 1 -  1_{\{  z_j =   z_l : j \in a, l \in b\}}  \bigg]   \\
			& = \chi_A(z_1,...,z_n)  \prod_{\substack{ \{  a, b \}    \in A :  \\ a \neq b}}  \bigg[ 1 -  1_{\{  z_j =   z_l : j \in a, l \in b\}}  \bigg]   \\
			& = \chi_A(z_1,...,z_n) \sum_{\sigma \in \{0,1\}^{A^{(2)}} } \prod_{ x \in A^{(2)}}  ( -  1_{\{  z_j =   z_l : j \in a, l \in b\}} )^{\sigma_x}   ,
		\end{align*} 
		where $A^{(2)}$ denotes the set of subsets of $A$ with two elements. 
		We note that  the map 
		\begin{align}  \label{bijlabel} \{0,1\}^{A^{(2)}} & \to \mathcal{P}(A^{(2)}), \\
			\sigma  & \mapsto \{ a : \sigma(a)=1\} \nn
		\end{align} 
		is  bijective (where $\mathcal{P}(M)$ denotes the power set of $M$, which 
		consists   of the subsets of $M$)
		and allows us to identity  $\{0,1\}^{A^{(2)}}$
		with subsets of  $A^{(2)}$, henceforth we shall identitify these sets. 
		We say that $A'$ is a coarser partition   than $A$ and write $A' \geq  A$  if for every $a \in A$ 
		there exists an $a' \in A'$ such that $a \subset a'$. 
		For $\sigma \in \{0,1\}^{A^{(2)}}$
		we define 
		$$A(\sigma) = \left\{ a \cup \bigcup_{b: \sigma\{a , b\}=1} b  : a \in A \right\} ,  $$
		which  is the partition which one obtains from $A$
		by forming unions of sets $a , b \in A$ for which $\sigma\{a,b\}=1$. 
		Now rearranging the sum we find 
		\begin{align} 
			\tilde{\chi}_A(z_1,...,z_n) &= \chi_A(z_1,...,z_n) \sum_{\sigma \in \{0,1\}^{A^{(2)}} } \prod_{ x \in A^{(2)}}  ( -  1_{\{  z_j =   z_l : j \in a, l \in b\}} )^{\sigma_x}  \nn \\
			& = \chi_A(z_1,...,z_n) 
			\sum_{A' : A' \geq A} \sum_{\sigma \in \{0,1\}^{A^{(2)}} } 1_{A(\sigma) = A'} \prod_{ x \in A^{(2)}}  ( -  1_{\{  z_j =   z_l : j \in a, l \in b\}} )^{\sigma_x} \nn  \\
			& = \sum_{A' : A' \geq   A}  \chi_{A'}(z_1,...,z_n)  
			\sum_{\substack{\sigma \in \{0,1\}^{A^{(2)}} \\  A(\sigma) = A' } } 
			( -  1 )^{\sum_{x \in A^{(2)}} \sigma_x}   \nn \\
			& = \sum_{A' : A' \geq   A}  \chi_{A'}(z_1,...,z_n)  
			\sum_{\substack{\sigma \subset {A^{(2)}} \\  A(\sigma) = A' } } 
			( -  1 )^{|\sigma|}  \label{eq:combinatorics} 
		\end{align} 
		where in the last line we made use of the bijection  \eqref{bijlabel}. 
		Using \eqref{upartition} we find 
		with 
		$
		M_n = {\bf E}\omega_x^n 
		$
		and \eqref{eq:combinatorics} 
		\begin{align*}
			{\bf E} ( \omega_{z_1 } \cdots \omega_{z_n}  ) 
			& = 
			\sum_{A \in \mathcal{A}_n} {\bf E} ( \omega_{z_1 } \cdots \omega_{z_n}   \tilde{\chi}_A(z_1,...,z_n) )\\
			& = 
			\sum_{A \in \mathcal{A}_n} \prod_{a \in A}  M_{|a|}   \tilde{\chi}_A(z_1,...,z_n) \\
			& = \sum_{A \in \mathcal{A}_n} \prod_{a \in A}  M_{|a|} \sum_{A' \in \mathcal{A}_n} 1_{A' \geq   A}  \chi_{A'}(z_1,...,z_n)  
			\sum_{\substack{\sigma \subset {A^{(2)}} \\  A(\sigma) = A' } } 
			( -  1 )^{|\sigma|} \\
			& = \sum_{A' \in \mathcal{A}_n} \chi_{A'}(z_1,...,z_n)  \left[ \sum_{\substack{ A \in \mathcal{A}_n \\ A' \geq   A }} \prod_{a \in A}  M_{|a|} 
			\sum_{\substack{\sigma \subset {A^{(2)}} \\  A(\sigma) = A' } } 
			( -  1 )^{|\sigma|} \right] 
		\end{align*}
		We define 
		\begin{align} 
			c_{B} := \sum_{\substack{ A \in \mathcal{A}_n \\ B \geq   A }} \prod_{a \in A}  M_{|a|} 
			\sum_{\substack{\sigma \subset {A^{(2)}} \\  A(\sigma) = B } } 
			( -  1 )^{|\sigma|} 
		\end{align} 
		and get 
		\begin{align}
			&  {\bf E} ( \omega_{z_1 } \cdots \omega_{z_n}  ) = \sum_{B  \in \mathcal{A}_n} c_B \chi_{B}(z_1,...,z_n) .
		\end{align}
		Using this identity we can calculate the expectations.
		Then 
		\begin{align} \label{exppatition}
			 {\bf E} \hat{V}(k_1) \cdots \hat{V}(k_n)
			& ={\bf E}  |\Lambda_L|^{-n} \sum_{x_1,...,x_n  \in \Lambda_L} \omega_{x_1}  e^{  i  k_1 \cdot x_1} \cdots 
			\omega_{x_n}  e^{  i  k_n \cdot x_n} \nn \\
			& =  |\Lambda_L|^{-n} \sum_{x_1,...,x_n  \in \Lambda_L} 
			\sum_{B  \in \mathcal{A}_n} c_B\chi_{B}(x_1,...,x_n)   e^{ i  k_1 \cdot x_1} \cdots 
			e^{  i  k_n \cdot x_n }  \nn \\
			& = |\Lambda_L|^{-n} 
			\sum_{B  \in \mathcal{A}_n} \sum_{b \in B} \sum_{x_b   \in \Lambda_L }      c_B  \prod_{b \in B} e^{ i  \sum_{j \in b} k_j \cdot x_b}  \nn \\
			&
			=  |\Lambda_L|^{-n} 
			\sum_{B  \in \mathcal{A}_n}       c_B|\Lambda_L|^{|B|}  \prod_{b \in B}\delta_{\Sigma_{j \in b} k_j ,0} 
		\end{align} 
		and we use for $k \in \widehat{\Lambda}_L$, that 
		\begin{align}\label{deltaident} 
			|\Lambda_L|^{-1} \sum_{x \in \Lambda_L} e^{- i k \cdot x }  = \delta_{k,0} .
		\end{align} 
		We expand the trace with respect to the eigenbasis 
		\begin{align*}\label{expbound-1-proof}
			&\mathbf{E}\,{\rm{tr}}[ F_1(H_0^L)V_L F_2(H_0^L)V_LF_3(H_0^L) \cdots V_L F_{n+1}(H_0^L)] \\
			&\quad= \mathbf{E}\sum_{k_0,k_1, \dots,k_n\in \hat{\Lambda}_L } \langle \varphi_{k_0} ,  F_{1}(H_0^L) \varphi_{k_0} \rangle 
			\langle  \varphi_{k_0}, V_L \varphi_{k_1}\rangle \langle \varphi_{k_1} F_{2}(H_0^L) \varphi_{k_1} \rangle  \nn\\
			&\qquad \qquad\cdots  \nn \langle  \varphi_{k_{n-1}} V_L \varphi_{k_n}\rangle \langle \varphi_{k_n} F_{n+1}(H_0^L) \varphi_{k_n} \rangle  \delta_{k_n,k_0}  .
		\end{align*} 
		Now evaluating the expression, by using that $\varphi_p$ is an eigenbasis of $H_0$, we find, 
		by means of the following change of variables
		\begin{align}
			q_0 = k_0 ,\quad  q_1 = k_1 - k_0 , \quad q_2 = k_2 - k_1 , \quad ...    \quad  q_n = k_n -k_{n-1} ,
		\end{align} 
		that
		\begin{align*}
			& k_0 = q_0 \\ 
			& k_j = \sum_{s=0}^j q_s  =: [\Sigma q]_j   \quad j=1,...,n .
		\end{align*} 
		Thus we can rewrite the trace as
		\begin{align}\label{expbound-1-proof}
			&\mathbf{E}\,{\rm{tr}}[ F_1(H_0^L)V_L F_2(H_0^L)V_LF_3(H_0^L) \cdots V_L F_{n+1}(H_0^L)] \\
			&\quad=  \mathbf{E}\sum_{k_0, k_1,...k_n\in \hat{\Lambda}_L }  
			F_{1}(\mathfrak{e}_L(k_0)) \hat{V}(k_1-k_0 )  F_{2}(\mathfrak{e}_L(k_1)) \hat{V}(k_2-k_1 ) \nn \\
			&\qquad \qquad\cdots  \nn \hat{V} (k_{n}-k_{n-1})F_{n+1}(\mathfrak{e}_L(k_n))) \delta_{k_n,k_0}  \\
			&\quad=  \mathbf{E}\sum_{q_0,q_1,...q_n\in \hat{\Lambda}_L }  
			F_{1}(\mathfrak{e}_L(q_0))  \hat{V}(q_1 )  F_{2}(\mathfrak{e}_L( [\Sigma q]_1) )  \hat{V}(q_2) \cdots \hat{V}(
			q_n  ) F_{n+1}(\mathfrak{e}_L([\Sigma q]_n ))   \delta_{[\Sigma q]_n ,q_0}  \nn 
		\end{align} 
		Taking the expectation we find  with  \eqref{exppatition}
		\begin{align} \label{expbound-1-proof0-0}
			&{ \bf E}  \sum_{q_0,q_1,...q_n\in \hat{\Lambda}_L }  
			F_{1}(\mathfrak{e}_L(q_0))  \hat{V}(q_1 )  F_{2}(\mathfrak{e}_L( [\Sigma q]_1) )  \hat{V}(q_2) \cdots \hat{V}(
			q_n  ) F_{n+1}(\mathfrak{e}_L([\Sigma q]_n ))  \delta_{[\Sigma q]_n ,q_0} \nn \\
			& =  |\Lambda_L|^{-n} \sum_{q_0,q_1,...q_n\in \hat{\Lambda}_L }  
			\sum_{B  \in \mathcal{A}_n}       c_B  |\Lambda_L|^{|B| }\prod_{b \in B}  \delta_{\Sigma_{j \in b} q_j ,0} \nn \\
			& \quad \times
			F_{1}(\mathfrak{e}_L(q_0))   F_{2}(\mathfrak{e}_L( [\Sigma q]_1) )  
			\cdots  F_{n+1}( \mathfrak{e}_L([\Sigma q]_n)]))  \delta_{[\Sigma q]_n ,q_0}
		\end{align} 	
		To make the notation more usable we define the set of all  indices, which are the maximum of a set $A \in \mathcal{A}_n$ by
		\begin{align}
			J_A := \{ \max a : a \in A \}
		\end{align}
		and the complement
		\begin{align}
			I_A :=  \{ 1 ,  \ldots ,  n \} \setminus J_A .
		\end{align}
		Since $A$ is a partition, for any $j \in \{1 , \ldots , n\}$ there is a unique set $a  \in A$ such that $j \in a$, we denote this set by $a(j)$. 
		We define the map 
		$M_A :( \hat{\Lambda}_L)^{I_A} \to (\hat{\Lambda}_L)^{n}$ as 
		\begin{align}
			[M_A (v)]_j :=
			\begin{cases}
				v_j , &  j \in I_A
				\\
				- \sum_{ l \in a(j) \setminus \{ j \}   } v_l ,  & j \in J_A ,
			\end{cases}
		\end{align}
		where $v=(v_j)_{j\in I_A}\in (\hat{\Lambda}_L)^{I_A}$. 
		It follows that 
		\begin{align}
			\sum_{j=1}^n[M_A(v)]_j  = 0 
		\end{align} 
		in $\hat{\Lambda}_L$. In fact, $M_A$ is a bijection from $(\hat{\Lambda}_L)^{I_A}$ onto
		\begin{align*}
			\{ q\in (\hat{\Lambda}_L)^n : \sum_{j\in a}q_j=0 \text{ for every } a\in A \}.
		\end{align*} 
		Now  integrating out the variables $q_j$ with $j \in J_A$  we obtain the following identity 
		\begin{align}\label{expbound-1-proof-3-4-5-9-0}
			\eqref{expbound-1-proof0-0} 
			& =   |\Lambda_L|^{-n}  \sum_{q_0 \in \hat{\Lambda}_L }
			\sum_{A  \in \mathcal{A}_n}  c_A  |\Lambda_L|^{|A| }
			\left(\prod_{s \in I_A} \sum_{ v_s  \in \hat{\Lambda}_L } \right)\\
			& \quad \times 
			F_{1,E}(\mathfrak{e}_L(q_0))   F_{2,E}(\mathfrak{e}_L( \sum_{j=1}^1[M_A(v)]_j  + q_0 ) ) 
			F_{3,E}(\mathfrak{e}_L( [\sum_{j=1}^2[M_A(v)]_j  + q_0 )) \nn
			\\ 
			& \quad 
			\quad \cdots F_{n+1,E}( \mathfrak{e}_L(\sum_{j=1}^n[M_A(v)]_j  + q_0 ) ) 
			\delta_{q_0,\sum_{j=1}^n[M_A(v)]_j  + q_0 }. \nn
		\end{align} 
		Note that $|I_A|=n-|A|$. Thus it can be written as 
		\begin{align}\label{expbound-1-proof-3-4-5-6-0}
			\eqref{expbound-1-proof0-0}  &=   
			\sum_{q_0 \in \hat{\Lambda}_L }
			\sum_{A  \in \mathcal{A}_n}  c_A  |\Lambda_L|^{-|I_A| }
			\left(\prod_{s \in I_A} \sum_{ v_s  \in \hat{\Lambda}_L }\right)  \\
			& \quad \times 
			F_{1,E}(\mathfrak{e}_L(q_0)) \left\{  \prod_{j=1}^{n}  F_{j+1,E}(\mathfrak{e}_L( \sum_{l=1}^j[M_A(v)]_l  + q_0 ))  
			\right\}   \nn
		\end{align} 
		We estimate  all  expressions $F_{j+1}$  for which $j \in J_A$ trivially by $\| F_{j+1} \|_\infty$.
		To estimate  the remaining terms, we shall use
		that the functions $\{  u_0 + \sum_{l=1}^{j}[M_A (u)]_l : j  \in I_A \} $ are linearly independent.
		More precisely, we have for $j \in I_A$ 
		\begin{align} \label{inductionrel-0} 
			\sum_{l=1}^{j}[M_A (v)]_l   = v_{j} +   \sum_{\substack{ l \in I_A, \,l < j\\ \max a(l)>j}}  v_l  .
		\end{align} 	
For a proof of  \eqref{inductionrel-0}, see reference \cite[Lemma 5.1]{HaslerKoberstein.2024-2}.

		By \eqref{inductionrel-0} we can use the following bijective change of variables between $(t_j)_{j \in I_A}$
		and  $(v_j)_{j \in I_A}$ 
		\begin{align}
			t_j  =\sum_{l=1}^j  [M_A(v)]_l + q_0 , \quad j \in I_A 
		\end{align} 
		Inserting this we find  estimating the terms 
		$ F_{j+1,E}(\mathfrak{e}_L( \sum_{l=1}^j[M_A(v)]_l  + q_0 ))  $ for $j \in J_A$ with $j < n$
		by the operator norm
		\begin{align}
			| \eqref{expbound-1-proof0-0}  |&=
			\bigg|\sum_{q_0 \in \hat{\Lambda}_L }
			\sum_{A  \in \mathcal{A}_n}  c_A  |\Lambda_L|^{-|I_A| }
			\left(\prod_{s \in I_A} \sum_{ v_s  \in \hat{\Lambda}_L } \right)	F_{1,E}(\mathfrak{e}_L(q_0))
			\nn \\
			& \quad \times 
		 \prod_{j \in I_A }  F_{j+1,E}(\mathfrak{e}_L( \sum_{l=1}^j[M_A(v)]_l  + q_0 ))  
			\prod_{j \in J_A }  F_{j+1,E}(\mathfrak{e}_L( \sum_{l=1}^j[M_A(v)]_l  + q_0 ))    \bigg | \nn \\
			&\leq
			\sum_{q_0 \in \hat{\Lambda}_L }
			\sum_{A  \in \mathcal{A}_n} | c_A|  |\Lambda_L|^{-|I_A| }
			\left(\prod_{s \in I_A} \sum_{ v_s  \in \hat{\Lambda}_L } \right)|F_{1,E}(\mathfrak{e}_L(q_0)) ||  F_{n+1,E}(\mathfrak{e}_L(q_0)) |
			\nn \\
			& \quad \times 
			\prod_{j \in I_A } | F_{j+1,E}(\mathfrak{e}_L( \sum_{l=1}^j[M_A(v)]_l  + q_0 ))  |
			\prod_{j \in J_A  : j < n } \|  F_{j+1,E} \|_\infty  \nn\\
			&= 
			\sum_{q_0 \in \hat{\Lambda}_L }
			\sum_{A  \in \mathcal{A}_n} | c_A | |\Lambda_L|^{-|I_A| }
			\left(\prod_{s \in I_A} \sum_{ t_s  \in \hat{\Lambda}_L } \right)|F_{1,E}(\mathfrak{e}_L(q_0)) F_{n+1,E}(\mathfrak{e}_L(q_0))  | 
			\nn \\
			& \quad \times 
			\prod_{j \in I_A } | F_{j+1,E}(\mathfrak{e}_L(t_s ))  |
			\prod_{j \in J_A  : j < n } \|  F_{j+1,E} \|_\infty.  \label{finalest}
		\end{align} 
		Thus we can use {\color{black}$|\Lambda_L|^{-|I_A|}$} to control the remaining sums. 
	\end{proof}

Let us now derive the following corollaries of Theorem \ref{main:graphbound}, where we will make use  of  \eqref{uppersemig}.

	\begin{corollary}\label{Contractionapp 1}
		Assume $\mathbf{E}[\omega_0]=0$,  and  let   $m \in \N$ with   $\mathbf{E}[\omega_0^{2m}]<\infty$. Let
$E$ be a regular point such that 
 \eqref{regularpoint1}  holds for $\delta_0 \in (0,1)$.
Then 
there exists a constant $C_m$ such that for all $\varepsilon \in (0,\delta_0]$ 
		\begin{align}
			\limsup_{L\to\infty}|\Lambda_L|^{-1}\mathbf{E} \| [V_L (H_0^L -E)^{-1}1_{\tilde I^c}(H_0^L) ]^m\|_2^2 \le C_m \varepsilon^{-m}(1+ |\ln \varepsilon|)^{2m} \label{n term} .
		\end{align}
	\end{corollary}
	\begin{proof}
		We can insert a identity in the middle and write the expectation as
		\begin{align*}
			\mathbf{E} \,{\rm tr} \,[G_1 V_L]^m \, {\rm Id} \,[V_L G_1]^m
		\end{align*}
		with $G_1=(H_0^L-E)^{-1}1_{\tilde I^c}(H_0^L)$. We  apply Theorem \ref{main:graphbound}  
for the functions $F_{j}(r) = (r- E)^{-1}1_{\tilde I^c}(r)$, if $j \in \{1,...,2m+1\} \setminus \{ m+1\}$, and $F_{m+1}(r) = 1$. 
We bound every $F_{j+1}$ with $j\in J_A\setminus \{2m\}$ trivially by $\varepsilon^{-1}$. This might include the identity. Here we get a factor of $\varepsilon^{-|J_A| +1} \le \varepsilon^{-m+1}$, since $|J_A|$ is at most $m$ by $\mathbf{E}[\omega_0]=0$. 
By taking the following upper semicontinuous majorant 
$|F_1(r)F_{2m+1}(r)| 
\leq |r-E|^{-2}1_{\overline{\tilde I^c}}(r)  $ 
we can estimate   the  averaged limit  superior  by \eqref{uppersemig},  and  get a factor $C\varepsilon^{-1}$ by integrating the majorant using \eqref{Int 1}. Similarly, we integrate the remaining functions using \eqref{Int 2} to get $|I_A|$ logarithmic contributions $C(1+|\ln\varepsilon|)$. Overall we arrive at 
		\begin{align*}
			\limsup_{L\to\infty}|\Lambda_L|^{-1}\mathbf{E} \| [V_L (H_0^L-E)^{-1}1_{\tilde I^c}(H_0^L) ]^m\|_2^2 \le C_m \varepsilon^{-m} (1+|\ln \varepsilon|)^{2m},
		\end{align*}
		where we bounded the number of logarithmic terms from above by $2m$.
	\end{proof}
	\begin{corollary}\label{Contractionapp 2}
		Assume   $\mathbf{E}[\omega_0]=0$, and let $k \in \N$ with $\mathbf{E}[\omega_0^{2k}]<\infty$. Let 
$E$ be a regular point such that 
 \eqref{regularpoint1}  holds for $\delta_0 \in (0,1)$.
Then 
there exists a constant $C_k$ such that for all $\varepsilon \in (0,\delta_0]$
		\begin{align}
			\limsup_{L\to\infty}|\Lambda_L|^{-1}\mathbf{E} \| 1_{\tilde I}(H_0^L) [V_L (H_0^L-E)^{-1}1_{\tilde I^c}(H_0^L) ]^k\|_2^2 \le C_k \varepsilon^{-k+1}(1+ |\ln \varepsilon|)^{2k} \label{k term}
		\end{align}
		\begin{align}
			\limsup_{L\to\infty}|\Lambda_L|^{-1}\mathbf{E} \left\| \frac{1_{\tilde I^c}(H_0^L)}{H_0^L-E}\left( V_L  \frac{ 1_{\tilde I^c}(H_0^L)}{H_0^L - E   }  \right)^k\right\|_2^2 \le C_k \varepsilon^{-k-1}(1+ |\ln \varepsilon|)^{2k} \label{k term for B}
		\end{align}
In case $k=1$, the logarithmic factors  \eqref{k term} and \eqref{k term for B} can be dropped. 
	\end{corollary}
	\begin{proof} First we show \eqref{k term}. 
		We write the expectation using cyclicity as
		\begin{align*}
			\mathbf{E} \,{\rm tr} \,1_{\tilde I}(H_0^L)[V_L G_1]^k [ G_1V_L]^k1_{\tilde I}(H_0^L) .
		\end{align*}
We  apply Theorem \ref{main:graphbound}  
for the functions $F_j(r) = 1_{\tilde I}(r)$, if $j \in \{ 1,2k+1\}$,    $F_{j}(r) = (r- E)^{-1}1_{\tilde I^c}(r)$, if $j \in \{2,...,2k\} \setminus \{ k+1\}$, and $F_{k+1}(r) = (r- E)^{-2}1_{\tilde I^c}(r)$.  The averaged limit superior of the sum of    $|F_1(\mathfrak{e}_L(q))F_{2k+1}(\mathfrak{e}_L(q))|= 1_{\tilde I}(\mathfrak{e}_L(q))$ over $q \in \hat{\Lambda}_L$ can be estimated by means of \eqref{uppersemig}  using \eqref{Int 3}, which yields  the  bound   $C\varepsilon$.
	For the remaining terms we can separate into the cases where $k$, the label  of   the  function $F_{k+1}$, falls either into $J_A\setminus\{2k\}$ or $I_A$. In the first case, we will get a bound $\prod_{j \in J_A \setminus \{ 2 k\} } \| F_{j+1} \|_\infty \leq \epsilon^{-|J_A|}$, by the supremum,  
 and  the summation over the remaining terms can be estimated by means of \eqref{Int 2}
resulting in an upper  bound $(1+ |\ln \varepsilon|)^{|I_A|}$.
In the second case, we will get a bound  $\prod_{j \in J_A \setminus \{ 2 k\} } \| F_{j+1} \|_\infty \leq \epsilon^{-|J_A|+1}$,
again by the supremum, 
 and  the summation over the remaining terms can be estimated by means of \eqref{Int 2} and \eqref{Int 1} 
resulting in an upper  bound $(1+ |\ln \varepsilon|)^{|I_A|-1} \varepsilon^{-1}$
In both cases we obtain the overall bound 
		\begin{align*}
			\limsup_{L\to\infty}|\Lambda_L|^{-1}\mathbf{E} \| 1_{\tilde I}(H_0^L) [V_L (H_0^L-E)^{-1}1_{\tilde I^c}(H_0^L) ]^k\|_2^2 \le C_k \varepsilon^{-k+1}(1+ |\ln \varepsilon|)^{2k},
		\end{align*}
		where we used that $|J_A|$ is at most $k$, since $\mathbf{E}[\omega_0] = 0$, 
and  we bounded the number of logarithmic terms from above by $2k$. This shows \eqref{k term}.

Estimate  \eqref{k term for B} is obtained analogously, except that to write    the absolute square
 we use  $F_j(r) = (r-E)^{-1} 1_{\tilde{I}^c}(r)$ if $j \in \{1,2k+1\}$, instead. 
Thus  to estimate the  sum of  
 $|F_1(\mathfrak{e}_L(q))F_{2k+1}(\mathfrak{e}_L(q))|= 1_{\tilde {I}^c}(\mathfrak{e}_L(q)) (\mathfrak{e}_L(q)-E)^{-2}$ over $q \in \hat{\Lambda}_L$
we use the following upper semicontinuous majorant 
$$
| 1_{\tilde {I}^c}(r) (r -E)^{-2} | \leq  | 1_{\overline{\tilde {I}^c}}(r) (r -E)^{-2} | 
$$
 and obtain by means of \eqref{uppersemig} and   \eqref{Int 1}  the  bound  $C\varepsilon^{-1}$. The other terms remain unchanged. \\
The case $k=1$ for \eqref{k term} , can be estimated  directly using  \eqref{Int 1}   and  \eqref{Int 3}.
 The case $k=1$ for \eqref{k term for B} , can be estimated  directly using  \eqref{Int 1}.
	\end{proof}

\section{Acknowledgements} 

D.H. wants to thank Gian Michele Graf for helpfull comments
and for   pointing out  the  Diploma thesis  \cite{AhlmDiploma}.  In addition 
D.H. wants to thank   Laszlo Erdoes for valuable discussions.

	\appendix
	
	\section{Wegner Estimate}\label{sec:wegner} 
	\begin{theorem} \label{thm:wegner} \cite{Wegner.1981}
			Assume $\{\omega_i\}_{i\in\mathbb{Z}^d}$ are i.i.d random variables with common bounded density $\rho$. For any $\lambda>0$ and any bounded interval $I$ the density of states measure satisfies the bound
			\begin{align}
				N_\lambda(I)  \le \frac{ \|\rho\|_\infty}{\lambda}|I|. \label{Wegner estimate}
			\end{align}
	\end{theorem}
	\begin{proof} 
 	Let $z=E+i\mu$ with $\mu>0$.
	Write the operator as
	\begin{align*}
		H_0+\lambda V = H_{\omega_0^\perp} +\lambda \omega_0  \delta_0  \langle\delta_0 , \cdot \rangle  
	\end{align*}
	where we defined
	\begin{align*}
		H_{\omega_0^\perp}= H_0+\lambda\sum_{x\ne0}\omega_x \delta_x   \langle\delta_x , \cdot \rangle .
	\end{align*}
	This operator is  selfadjoint and independent of $\omega_0$. We apply the rank one formula, see e.g.  \cite[(I.13)]{R1},
	\begin{align*}
		\langle \delta_0, (H_\lambda-z)^{-1} \delta_0\rangle = \frac{m(z)}{1+ \lambda \omega_0 m(z)}= \frac{1}{\lambda \omega_0+m(z)^{-1}}
	\end{align*}
	with 
	\begin{align*}
		m(z) = \langle \delta_0, (H_{\omega_0^\perp}-z)^{-1} \delta_0\rangle.
	\end{align*}
	Since $H_{\omega_0^\perp}$ is selfadjoint, $m(z)$ is a Herglotz function. Thus for $\mathrm{Im}z>0$ we have $\mathrm{Im}\,m(z)^{-1}<0$. Now take expectation, $\mathbf{E}_0$, with respect to $\omega_0$. We estimate using the non negativity of the integrand
	\begin{align*}
	\mathbf{E}_0\,	\mathrm{Im}\langle \delta_0, (H_\lambda-z)^{-1} \delta_0\rangle \le \|\rho\|_\infty \int \mathrm{Im}\frac{1}{\lambda \omega_0+m(z)^{-1}} d\omega_0 =\pi \,\frac{ \|\rho\|_\infty}{\lambda}
	\end{align*}
	where we used $\int \mathrm{Im} (u-z)^{-1} du=\pi$ for $\mathrm{Im}z >0$. Averaging over the other random variables leaves the bound unchanged.
	Thus it follows
	\begin{align*}
		\frac{1}{\pi}\mathbf{E}\,	\mathrm{Im}\langle \delta_0, (H_\lambda-z)^{-1} \delta_0\rangle  \le \frac{ \|\rho\|_\infty}{\lambda}
	\end{align*}
	Now the statement follows by integrating the expression and using Stone's formula, cf.  \cite{SR1}, by letting $\mu\downarrow 0$.
\end{proof}

	\section{Integral Bounds} \label{sec:integralbounds} 
	Assume $E$ is a regular point of $H_0$ with respect to some $\delta_0\in(0,1)$, that is,
	\begin{align*}
		N_0([E-\delta,E+\delta])\le \Gamma(E)\delta
	\end{align*}
	for all $\delta\in[0,\delta_0]$. Then the following  bounds hold for all $\varepsilon\in(0,\delta_0]$
	\begin{align}
		&\int_{\{|\mathfrak{e}(s)-E|\ge \varepsilon\}}(\mathfrak{e}(s)-E)^{-2}\mathrm{d}\widehat{s} \leq  C_{2,E} \,\varepsilon^{-1}, \label{Int 1} \\
		&\int_{\{|\mathfrak{e}(s)-E|\ge \varepsilon\}}|\mathfrak{e}(s)-E|^{-1}\mathrm{d}\widehat{s} \leq  C_{1,E} \,(1+|\ln \varepsilon|), \label{Int 2} \\
		&\int_{\T^d}1_{[E-\varepsilon, E+\varepsilon]}(\mathfrak{e}(s))\mathrm{d}\widehat{s} \leq C_E\, \varepsilon. \label{Int 3}
	\end{align}
The inequalities  \eqref{Int 1}, \eqref{Int 2}, \eqref{Int 3}, follow from the more general statements
\eqref{Int a 1}, \eqref{Int a 2}, \eqref{Int a 3}, below. 
Assume, more generally, that  $E$ is a regular point of order $\alpha \in (0,1]$ for $H_0$ with respect to some $\delta_0\in(0,1)$, that is,
	\begin{align*}
		N_0([E-\delta,E+\delta])\le \Gamma(E,\alpha)\delta^\alpha
	\end{align*}
	for all $\delta\in[0,\delta_0]$, cf.  Definition \ref{def:regularpoint}. Then for $\kappa = 1,2$ the following  bound hold for all $\varepsilon\in(0,\delta_0]$
	\begin{align}
		&\int_{\{|\mathfrak{e}(s)-E|\ge \varepsilon\}}(\mathfrak{e}(s)-E)^{-\kappa}\mathrm{d}\widehat{s} \leq 
C_{\kappa,E,\alpha}\varepsilon^{\alpha-\kappa}, \quad \alpha < \kappa \label{Int a 1}  \\
		&\int_{\{|\mathfrak{e}(s)-E|\ge \varepsilon\}}|\mathfrak{e}(s)-E|^{-\kappa}\mathrm{d}\widehat{s} \leq
 C_{\kappa,E,\alpha} \,(1+|\ln \varepsilon|) , \quad \alpha = \kappa  \label{Int a 2} \\
		&\int_{\T^d}1_{[E-\varepsilon, E+\varepsilon]}(\mathfrak{e}(s))\mathrm{d}\widehat{s} \leq C_{E,\alpha}\, \varepsilon^\alpha. \label{Int a 3}
	\end{align}
		\begin{proof}
		The bound \eqref{Int a 3} follows directly by definition. Thus it remains to show  \eqref{Int a 1} and  \eqref{Int a 2}. Since the measure $N_0$ is the pushforward under the function $\mathfrak{e}$,  cf. \eqref{freedosispushforwarde}, we can write 
		\begin{align*}
			I:=\int_{\{|\mathfrak{e}(s)-E|\ge \varepsilon\}}|\mathfrak{e}(s)-E|^{-\kappa}\mathrm{d}\widehat{s} = \int_{|x-E|\ge \varepsilon}\frac{1}{|x-E|^\kappa}dN_0(x)
		\end{align*}
		Now use the Layer Cake representation
		\begin{align*}
			I= \kappa \int_\varepsilon^\infty N_0(\{\varepsilon\le |x-E|\le r\})r^{-1-\kappa}dr.
		\end{align*}
		We separate the domain into
		\begin{align*}
			I= \kappa \int_\varepsilon^{\delta_0} N_0(\{\varepsilon\le |x-E|\le r\})r^{-1-\kappa}dr+\kappa \int_{\delta_0}^\infty N_0(\{\varepsilon\le|x-E|\le r\})r^{-1-\kappa}dr
		\end{align*}
		For the second term we can simply use that the measure $N_0$ is bounded by 1 and integrate, to get a factor of $\delta_0^{-\kappa}$. For the first we use $N_0(\{\varepsilon\le|x-E|\le r\})\le N_0(\{|x-E|\le r\})\le \Gamma(E,\alpha)r^\alpha$ and integrate to get 

\begin{align*}
 \frac{\kappa}{\kappa-\alpha}\Gamma(E,\alpha)(\varepsilon^{\alpha-\kappa}-\delta_0^{\alpha-\kappa}) &  ,  \quad \text{if } \quad \alpha < \kappa , \\
   {\kappa}\Gamma(E,\alpha)( \ln\delta_0  - \ln \varepsilon  )  &  ,  \quad  \text{if } \quad \alpha = \kappa  .
\end{align*} 
Overall we get
		\begin{align*}
			I\le \left\{ \begin{array}{ll}  
 \frac{\kappa}{\kappa-\alpha}\Gamma(E,\alpha)(\varepsilon^{\alpha-\kappa}-\delta_0^{\alpha-\kappa})  +
 \kappa \delta_0^{-\kappa} \le C_{\kappa,E,\alpha}\varepsilon^{\alpha-\kappa},    &  \alpha < \kappa , \\ 
   {\kappa}\Gamma(E,\alpha)( \ln\delta_0  - \ln \varepsilon  )  +\kappa \delta_0^{-\kappa} \leq    C_{\kappa,E,\alpha} \,(1+|\ln \varepsilon|), & \alpha = \kappa . 
\end{array} \right. 
		\end{align*}
	This shows the claimed bounds. 
	\end{proof}
		\printbibliography
	
\end{document}